\documentclass[a4paper,11pt]{article}
\usepackage{jcappub} 

\arxivnumber{2304.07321}
\title{Constraints on the proton fraction of cosmic rays at the highest energies and the consequences for cosmogenic neutrinos and photons}

\usepackage{graphicx}
\usepackage{commath}
\usepackage{hyperref}
\usepackage{orcidlink}
\usepackage{multicol}
\usepackage[separate-uncertainty=true]{siunitx}
\DeclareSIUnit\parsec{pc}
\DeclareSIUnit\year{yr}
\DeclareSIUnit\erg{erg}
\usepackage[capitalise]{cleveref}
\crefname{figure}{Figure}{Figures}
\crefname{section}{Section}{Sections}
\crefname{table}{Table}{Tables}
\crefname{equation}{Eq.}{Eqs.}

\definecolor{ntnu}{RGB}{0, 80, 158}
\hypersetup{
   citecolor  = ntnu,
   urlcolor   = ntnu,
   linkcolor  = ntnu,
}

\graphicspath{{./}{images/}}
\newcommand{\rmax}{\ensuremath{R_\text{max}}}
\newcommand{\emax}{\ensuremath{E_\text{max}}}

\newcommand{\ptwo}{\text{PP}}
\newcommand{\mxmax}{\langle X_\text{max}\rangle}
\newcommand{\sxmax}{\sigma(X_\text{max})}

\author[a,b]{Domenik Ehlert \orcidlink{0000-0002-4322-6400}}
\emailAdd{domenik.ehlert@ntnu.no}
\author[c,b]{Arjen van Vliet \orcidlink{0000-0003-2827-3361}}
\author[a]{Foteini Oikonomou \orcidlink{0000-0002-0525-3758}}
\author[b]{Walter Winter \orcidlink{0000-0001-7062-0289}}

\affiliation[a]{Norwegian University of Science and Technology (NTNU), 7491 Trondheim, Norway}
\affiliation[b]{Deutsches Elektronen-Synchrotron DESY, Platanenallee 6, 15738 Zeuthen, Germany}
\affiliation[c]{Department of Physics, Khalifa University, P.O. Box 127788, Abu Dhabi, United Arab Emirates}

\abstract{
Over the last decade, observations have shown that the mean mass of ultra-high-energy cosmic rays (UHECRs) increases progressively toward the highest energies. However, the precise composition is still unknown and several theoretical studies hint at the existence of a subdominant proton component up to the highest energies.
Motivated by the exciting prospect of performing charged-particle astronomy with ultra-high-energy (UHE) protons we quantify the level of UHE-proton flux that is compatible with present multimessenger observations and the associated fluxes of neutral messengers produced in the interactions of the protons.
We study this scenario with numerical simulations of two independent populations of extragalactic sources and perform a fit to the combined UHECR energy spectrum and composition observables, constrained by diffuse gamma-ray and neutrino observations.
We find that up to of order $10\%$ of the cosmic rays at the highest energies can be UHE protons, although the result depends critically on the selected hadronic interaction model for the air showers. Depending on the maximum proton energy ($E_\text{max}^\text{p}$) and the redshift evolution of sources, the associated flux of cosmogenic neutrinos and UHE gamma rays can significantly exceed the multimessenger signal of the mixed-mass cosmic rays. Moreover, if $E_\text{max}^\text{p}$ is above the GZK limit, we predict a large flux of UHE neutrinos above EeV energies that is absent in alternate scenarios for the origin of UHECRs. We present the implications and opportunities afforded by these UHE proton, neutrino and photon fluxes for future multimessenger observations.
}

\begin{document}
\maketitle
\flushbottom

\section{Introduction}

Ultra-high-energy cosmic rays (UHECRs), charged particles of astrophysical origin with energy above $\sim10^{18}\,\si{\electronvolt}$, are the most energetic cosmic messengers and, as such, probes of the most extreme astrophysical environments. Because of extragalactic and Galactic magnetic fields, their sources remain elusive, even after years of high-precision observation by the latest generation of UHECR detectors, in particular the Pierre Auger Observatory (Auger) and the Telescope Array (TA).
    
Observations suggest that the composition of UHECRs is surprisingly pure, with each accelerated nuclear species only dominant in a very narrow band of the UHECR spectrum, and the entire spectrum is produced through a carefully tuned combination of the individual peaks (e.g.~\cite{Unger:2015laa,Aab:2016zth,AlvesBatista:2018zui,Heinze:2019jou}). The combination of a smooth increase of the average mass and pure composition at all energies implies that the population variance of sources must be remarkably low~(\cite{Ehlert:2022jmy}; see also \cite{Heinze:2020zqb}). Under these circumstances, the observed flux cutoff at $E_\text{CR}\gtrsim\SI{50}{\exa\electronvolt}$ is generally predicted to be an effect of the maximum particle energy reachable at the cosmic accelerators. Within this ``Peters cycle''~\cite{Peters:1961,Gaisser:2016uoy} model of cosmic-ray acceleration with rigidity-dependent maximum energy, no light cosmic rays (CRs) are expected at the highest energies.
    
Nevertheless, the existence of protons or light nuclei at the highest energies, where there are no measurements of composition-sensitive observables with the fluorescence detectors of Auger and TA, cannot be ruled out at present. A very interesting possibility would be the existence of an additional proton-dominated component at the highest energies.
Such a flux cannot be easily explained by reprocessing of accelerated UHECR within the source as proposed for extragalactic protons below the ankle, see e.g.~\cite{Unger:2015laa}, but must originate from a secondary population of independent sources that exclusively accelerates protons to ultra-high energies or where heavier nuclei are efficiently disintegrated before escaping the source region. Motivations for an additional source population come from the expected differences between possible UHECR accelerators, e.g. active galactic nuclei~\cite{Rodrigues:2020pli} or gamma-ray bursts~\cite{Waxman:1995vg}. Such a proton flux does not need to be produced by astrophysical processes necessarily, but could also originate from the decay of heavy dark matter~(e.g.\ \cite{Ishiwata:2019aet,Das:2023wtk}).
    
Circumstantial evidence for an additional proton component is provided by an apparent flattening of the increase in observed UHECR mass at $E_\text{CR}\gtrsim\,\SI{30}{\exa\electronvolt}$, as reported in an analysis of Auger surface detector data \cite{PierreAuger:2017tlx,ToderoPeixoto:2019bt}. This feature could indicate a flux of UHE protons with different spectral index to the bulk of the UHECRs, either from a secondary source population or from a single nearby source~\cite{Plotko:2022urd}, but it could also originate from a natural mass limit of the mixed UHECR flux.

Similar two-component models have been previously studied, either in the context of the transition region between Galactic and extragalactic cosmic rays below $10^{18.7}\, \si{\electronvolt}$~\cite{Mollerach:2020mhr,PierreAuger:2021mmt,Luce:2022awd,PierreAuger:2022atd}, or similar to the present paper at the highest energies~\cite{Muzio:2019leu,Das:2020nvx}.

Compared to Ref.~\cite{Das:2020nvx} we consider a much wider range of proton source parameters. In particular, we study three distinct scenarios for the spectrum injected by the UHE proton sources, of which only one is considered in \cite{Das:2020nvx}. We also take into account the production of cosmogenic photons, both in the GeV-TeV band and at ultra-high energy, and analyse quantitatively the impact of multimessenger constraints on the UHECR source parameters.

While we were finalising this article, another study by Muzio et al.\ \cite{Muzio:2023skc} on a subdominant population of UHE proton sources appeared. Unlike our work, they consider only mono-elemental injection by medium/heavy-composition cosmic ray sources and include in-source photohadronic interactions (see also \cite{Muzio:2019leu} for an earlier work using the same model). They assume a particular blackbody-like photon field within the source region resulting in photohadronic interactions of the cosmic rays, whereas we do not consider interactions in the source environment. Our results are therefore more general but the inferred parameters should be understood as effective parameters of the cosmic rays after they escape the source environment. In the case of astrophysical environments with small optical depth to hadronic and photohadronic interactions, such as low-luminosity gamma-ray bursts, BL Lac objects, and radio lobes of jetted active galactic nuclei, our results closely resemble the spectra produced inside the astrophysical sources. In addition, \cite{Muzio:2023skc} assume the same redshift evolution of the emissivity for both UHECR source populations and only optimise the contributions of both source populations sequentially. Our results are complementary and more general in terms of the source parameters of both source populations which we optimise simultaneously.
    
UHE protons, should they exist, are of significant interest for ``UHECR astronomy'' due to their high rigidity and consequently weak deflections in magnetic fields. Additionally, if they are accelerated to energies beyond $\sim10^{19.7}\si{\electronvolt}$, the cross-section for photo-pion production on CMB photons is enhanced due to the $\Delta$-resonance. This effect, known as the Greisen-Zatsepin-Kuzmin (GZK) limit~\cite{Greisen:1966jv,Zatsepin:1966jv}, leads to strong attenuation of UHE protons above this energy if they are produced in sources more distant than $\sim\SI{100}{\mega\parsec}$ (see e.g.\ \cite{Gaisser:2016uoy}) and the abundant production of charged and neutral pions. The subsequent decay of these pions will result in a large flux of high-energy neutrinos and gamma rays.

In this paper we quantify the maximum flux of UHE protons compatible with current observations of UHECR spectrum and composition, considering multimessenger constraints from gamma rays and neutrinos. We investigate two separate scenarios for the maximum proton energy; (i) a high-$\emax^\text{p}$ and (ii) a low-$\emax^\text{p}$ scenario.
    
A brief overview of the model is provided in \cref{sec:methods}. Injection and propagation of the cosmic rays are simulated with the Monte-Carlo framework \textsc{CRPropa\,3}~\cite{Batista:2016yrx,AlvesBatista:2022vem}, taking into account the interaction with the cosmic microwave background and extragalactic background light~\cite{Gilmore:2012}. The best-fit source parameters are obtained in \cref{sec:results_UHECR} by comparing the model predictions with existing observations, and in \cref{sec:results_multimessenger} we discuss the expected multimessenger signal. A specific, exotic scenario with flux recovery beyond the GZK cutoff is presented in \cref{sec:results_recovery}. Finally, we discuss our results in the context of similar existing studies in \cref{sec:paper_comparison}, and conclude in \cref{sec:conclusions} that current UHECR data is compatible with a significant contribution by this additional proton component of up to $15\%$ at \SI{20}{\exa\electronvolt}. The precise value depends critically on the choice of the hadronic interaction model for air shower modelling and the maximum proton energy.

\section{Methods}\label{sec:methods}
The primary, \textbf{mixed-composition, UHECR sources (MIX)} are modelled following the effective parametrisation introduced in \cite{Aab:2016zth} but with minor modifications detailed in \cite{Ehlert:2022jmy}. We assume the acceleration to be universal in particle rigidity\footnote{The rigidity of a particle is defined as $R=E/Z$ in natural units, where $Z$ is the nuclear charge. It is a measure of the susceptibility to magnetic deflections.}, following a ``Peters cycle'', with a power-law source spectrum and an exponential cutoff at the highest energies. Sources within the MIX population are assumed as identical, with a volumetric emission rate
\begin{equation}
    Q_{A}(E) = Q_{A}^{E_0}\,\left(\frac{E}{E_0}\right)^{-\gamma}\,\exp{\left(-\frac{E}{Z\,\emax^\text{p}}\right)}
\end{equation}
for the five injected elements $A\in$\{$^1$H, $^4$He, $^{14}$N, $^{28}$Si, $^{56}$Fe\}. Here $Q_{A}^{E_0}$ is the local $(z=0)$ emission rate at a normalisation energy $E_0\ll \emax^\text{p}$ in $\rm erg^{-1}~Mpc^{-3}~yr^{-1}$, and $\gamma$ is the spectral index which is $\approx 2$ for diffusive shock acceleration. The source emissivity, i.e.\ the luminosity density, can be derived from the emission rate as
\begin{equation}
    L_0 = \sum_A \int_{E_\text{min}}^{\infty}\dif E\,\left(E\cdot Q_A(E)\right)\,,
\end{equation}
where we have chosen $E_\text{min} = 10^{17.8}\,\si{\electronvolt}$.
    
The predicted flux at Earth for an observed nuclear mass $A'$ and energy $E'$, and for a redshift evolution $n(z)$ of the source population emissivity $Q_{A}(E)$, is
\begin{equation}
    \phi(E',A') = \sum_A\int\dif E\int\dif z \left|\frac{\dif t}{\dif z}\right|\,n(z) \,Q_A(E) \cdot\frac{\dif N_{A'}}{\dif E' \dif N_A}(E',E,z)\,.
\end{equation}
The last term translates the injected spectrum at the sources to the observed spectrum after propagation and is obtained via Monte-Carlo simulations with {\scshape CRPropa}.
In general, the population-emissivity redshift evolution $n(z)$ is composed of the evolution of per-source luminosities and the density evolution of the source population. In our analysis, we do not attempt to distinguish the difference between these effects and describe the evolution with a (broken) power law
\begin{equation}\label{eq:dndz}
    n(z) =
    \begin{cases}
	(1+z)^m &\text{for }m\leq0, \\
	(1+z)^m &\text{for }m>0\text{ and }z<z_0,\\
	(1+z_0)^m &\text{for }m>0\text{ and }  z_0 < z < z_\text{max},\\
	0 & \text{otherwise,}
    \end{cases}
\end{equation}
with $z_0=1.5$ and $z_{\max}=4$~\cite{vanVliet:2019nse}. Sources at $z\gtrsim1$ have a negligible impact on the observed UHECR flux because of attenuation effects, but they play an important role for the expected multimessenger signal of co-produced neutrinos and low-energy gamma rays. A more conservative estimate of the cosmogenic neutrino flux is obtained if these high-redshift sources, which cannot be constrained by the cosmic-ray fit, are ignored.
    
For the additional population of {\bf UHE pure-proton sources (PP)}, we are particularly interested in the predicted flux of cosmogenic neutrinos at $E_\nu\approx\SI{1}{\exa\electronvolt}$ since this corresponds to the peak sensitivity interval of many existing and planned neutrino experiments. If these neutrinos are produced in the interactions of cosmic rays with photon fields, they typically receive $\sim5\%$ of the primary CR energy~\cite{Gaisser:2016uoy}, which implies that the relevant energy is $E_\text{CR}\approx\SI{20}{\exa\electronvolt}$. We define this value as the reference energy at which we evaluate the contribution of the PP UHE protons to the observed flux of UHECRs. Properties of the pure-proton sources are described by the independent set of parameters $\emax^\ptwo,\,\gamma^\ptwo,\,m^\ptwo,{\rm and}\,L_0^\ptwo$.
    
The interactions of UHECRs with  cosmic background photons lead to the production of secondary photons and neutrinos, with the strength of this ``cosmogenic'' multimessenger signal depending predominantly on the cosmic-ray composition, injection spectral index and source distance. We compare our model predictions for the UHECR spectrum and composition with publicly available data by Auger~\cite{PierreAuger:2020qqz,Yushkov:2020nhr}. Since the composition cannot be observed directly, the mean, $\mxmax$, and standard deviation, $\sxmax$, of the depth of the air-shower maximum are used as proxy observables, and the conversion is performed with the hadronic interaction models \textsc{Epos-LHC}~\cite{Pierog:2015} and \textsc{Sibyll2.3c}~\cite{Fedynitch:2018cbl}. To minimise the impact of a possible contribution of other sources dominating the observed flux below the ankle we limit our analysis to $E_\text{CR}\geq10^{18.7}\,\si{\electronvolt}$. However, spectral points at lower energies are included as upper limits and scenarios with excessive sub-ankle flux are rejected.

The best-fit source parameters are determined in a two-step fitting process. We discretise the parameter space in maximum energy/rigidity, spectral index and redshift evolution for both source classes and sample a large number of possible combinations of these parameters. For each of these possible source configurations, we then use the Levenberg-Marquardt algorithm\footnote{Implemented in the \texttt{curve\_fit} routine from the \texttt{SciPy.optimize} library.} to find the injection fractions $f_A$ of the MIX sources and emissivities $[L_0,L_0^\ptwo]$ of both source populations that minimise the $\chi^2$ differences between our model predictions for the UHECR spectrum and composition and the Auger data points. If a reasonable fit ($\chi^2<250$) is found for a particular combination  of source parameters then adjacent points are also evaluated in an iterative process. This is different to the two-step approach of ~\cite{Muzio:2023skc} since we optimise the contributions of both source populations at the same time.

Constraints on the source parameters derived from a comparison of the predicted cosmogenic flux of gamma rays and neutrinos with observations and upper limits are taken into account with additional $\Delta\chi^2$-penalty terms. For observed fluxes, such as the Fermi-LAT IGRB~\cite{Fermi-LAT:2014ryh} and parts of the IceCube HESE neutrino flux~\cite{IceCube:2020wum} we consider a simple one-sided $\chi^2$ penalty that only contributes if the predicted flux exceeds observations. For upper-limit points with a low number of, or zero, events per bin, e.g.\ the Auger UHE neutrino~\cite{Pedreira:2021gcl} and UHE gamma-ray limits~\cite{Savina:2021cva,PierreAuger:2022aty} we use the Poisson likelihood $\chi^2$~\cite{Baker:1983tu} but the penalty is only applied if the predicted number of events in a bin exceeds the observed number. The relevant data sets are:
\begin{align}
        \Delta\chi^2_\nu:~&\text{IceCube HESE flux~\cite{IceCube:2020wum}}~\&~\text{Auger UHE neutrino limit~\cite{Pedreira:2021gcl}},  \\
        \Delta\chi^2_\gamma:~&\text{Fermi-LAT IGRB flux~\cite{Fermi-LAT:2014ryh}}\,,~\text{Auger hybrid UHE gamma-ray limit~\cite{Savina:2021cva}}   \nonumber \\
        &\&~\text{Auger SD UHE gamma-ray limit~\cite{PierreAuger:2022aty}}.   \nonumber
\end{align}
We exclude possible source configurations where the combined multimessenger penalty exceeds the level of two sigma, i.e.\ when $\Delta\chi^2_\nu+\Delta\chi^2_\gamma > 4$. However, in the plots of the cosmogenic neutrino and gamma-ray fluxes we only include the rejections by the respective messenger.

\section{Fit with an Additional Proton Component}\label{sec:results_UHECR}
\begin{figure}
    \centering
    \includegraphics[width=0.48\linewidth]{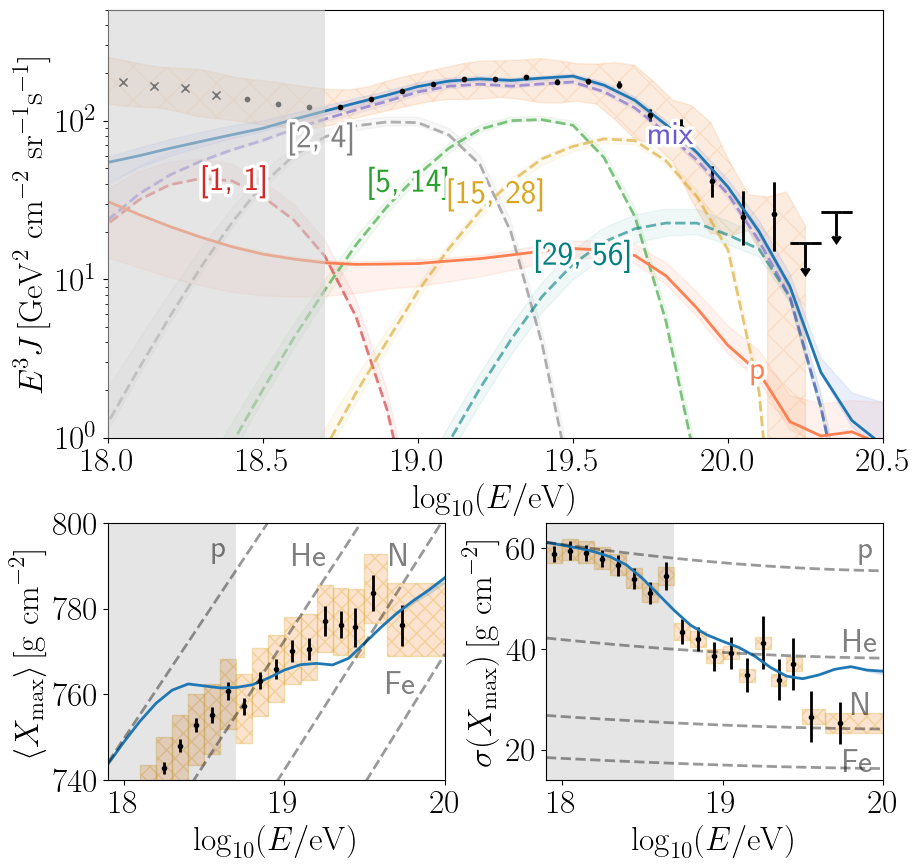}
    \includegraphics[width=0.48\linewidth]{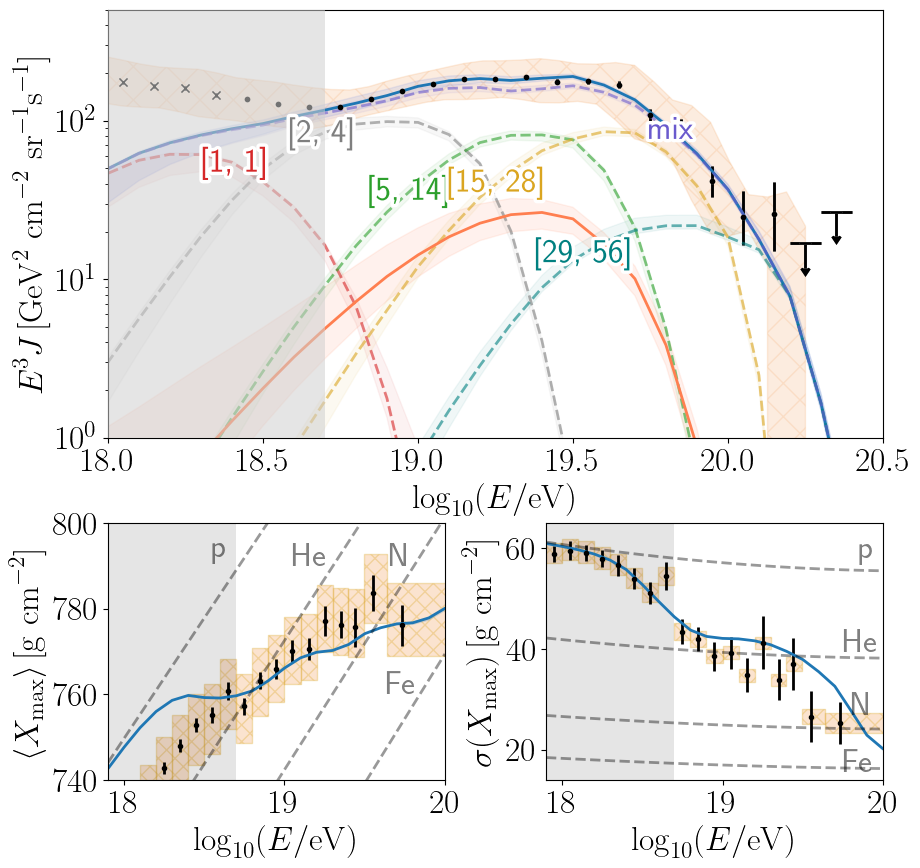}
    \caption{Predicted spectrum and composition at Earth for the investigated scenarios, with {\scshape Epos-LHC} as hadronic interaction model. Left: ``proton-dip'' (2SC-dip). Right: ``UHECR'' best fit (2SC-uhecr). Best-fit parameter values are listed in~\cref{tab:results_epos}. Dashed lines indicate the contributions of the separate mass groups from the mixed-composition sources, with $[A_\text{min},A_\text{max}]$. The additional protons from the second population are shown as a solid, orange line. Coloured bands indicate the $68\%$ uncertainties.}\label{fig:EPOS_2SC_CR}
\end{figure}
	
We investigate two different scenarios in terms of the proton maximum energy, assuming {\scshape Epos-LHC} as hadronic interaction model. Results for {\scshape Sibyll2.3c} are shown in Appendix \ref{apx:sibyll_fit}. In both scenarios, we find the redshift evolution of the PP number density to be unconstrained by cosmic-ray observations alone. Since the PP flux is pure protons, interactions during propagation do not affect the observed composition. However, propagation effects soften the distribution and attenuate the original UHE proton flux. Stronger redshift evolutions require harder injection spectra and higher source emissivity.

The two models described in the following, with maximum energy of the protons fixed to $10^{5}\,\si{\exa\electronvolt}$ and $\SI{10}{\exa\electronvolt}$ respectively, represent the most characteristic scenarios identified during a scan of $\emax^\ptwo$ (see \cref{apx:proton_emax}).
	
\subsection{Two-Source-Class Dip Model (2SC-dip)}
We are particularly interested in scenarios that produce a large flux of UHE neutrinos and gamma rays. This requires proton energies sufficiently above the GZK limit to enable copious photo-pion production on CMB photons, and we, therefore, choose $\emax^\ptwo=10^{23}\,\si{\electronvolt}$ for our first scenario. The best-fit properties of both source populations are listed in \cref{tab:results_epos}, 2nd column, and the predicted spectrum and composition at Earth are shown in \cref{fig:EPOS_2SC_CR}, left. The preferred maximum rigidity, spectral index and redshift evolution of the mixed-composition source population are compatible with the values obtained for the single-population model within uncertainties, and the additional protons provide a relatively constant contribution of approx.\ $5-10\%$ between the ankle and the end of the GZK cutoff (\cref{fig:2SC_proton_fraction}, teal band). We find that the overall shape effectively corresponds to the predictions from the classical ``proton-dip'' explanation of the UHECR flux~\cite{Berezinsky:1988wi}. While this model is inconsistent with current measurements of the UHECR composition and the high-energy neutrino flux~\cite{Heinze:2015hhp}, our results show that it can still be relevant if the total proton contribution remains subdominant to the primary, mixed-composition, cosmic-ray flux. We refer to the presented source model as the ``dip'' or 2SC-dip (two-source-class dip) model.
 
Here, the proton sources are required to exhibit a soft injection spectrum (see \cref{apx:proton_parspace}), which could be a distinguishing feature of this additional source population in the observed flux, provided that reliable event-by-event mass reconstruction becomes available in the future. Softer spectra than suggested by the best fit are disfavoured since the associated sub-ankle flux would exceed observational limits. For hard spectra, $\gamma^\ptwo\lesssim2$, the additional protons only contribute at energies around the GZK cutoff and the possibilities for improving the fit over the entire energy range are consequently limited. The combination of both effects results in a clearly localised preferred spectral index of the proton sources. 
	
\subsection{Two-Source-Class Best-Fit Model (2SC-uhecr)}
\begin{figure}
    \centering
    \includegraphics[width=0.7\linewidth]{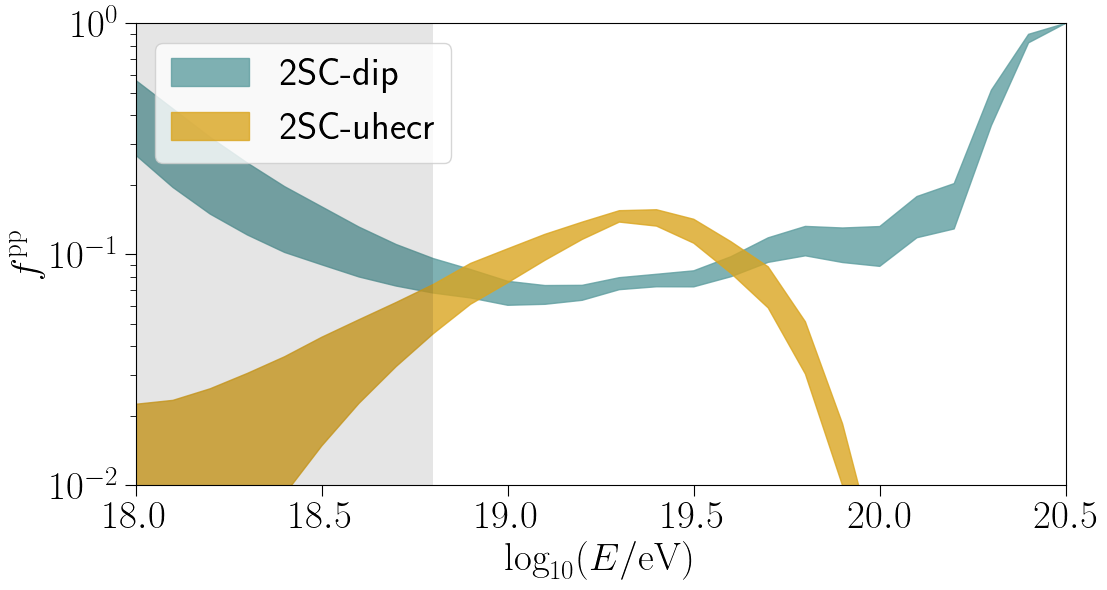}
    \caption{Contribution of the PP protons to the observed, differential UHECR flux as a function of energy, within $1\sigma$ of the best fit to CR spectrum and composition (see \cref{fig:EPOS_2SC_CR}).}\label{fig:2SC_proton_fraction}
\end{figure}

An alternative scenario is presented by proton sources with energies comparable to the standard, mixed-composition, cosmic-ray sources. For this model, we set $\emax^\ptwo=\,\SI{10}{\exa\electronvolt}$. At the best fit (\cref{tab:results_epos}, 3rd column), the improvement over the dip-model is $\Delta\chi^2\approx-15$ but very hard proton spectra are required (see \cref{apx:proton_parspace}). The predicted PP proton spectrum at Earth exhibits a peak-like shape reminiscent of the individual, peaked, mass groups originating from the mixed composition sources (\cref{fig:EPOS_2SC_CR}, right). However, due to the choice of $\emax^\ptwo$, the peak energy is shifted upward by approximately an order of magnitude compared to the mixed-population proton peak. Compared to the 2SC-dip model, the best-fit observed proton fraction at \SI{20}{\exa\electronvolt} is significantly larger, up to $15\%$, but the contribution is limited to a small energy interval and becomes negligible below the ankle (\cref{fig:2SC_proton_fraction}, brown band).
	
While this scenario, the ``UHECR best-fit'' model (2SC-uhecr), provides a significant improvement in the cosmic-ray fit, it comes at the cost of extremely hard proton injection spectra, and the expected cosmogenic neutrino and UHE gamma-ray signal associated with the protons is reduced due to the sub-GZK maximum proton energies. With the injection spectrum of the additional protons similar to the bulk of the cosmic rays, separation of the two components will be difficult even if event-by-event mass reconstruction were available. However, the predicted existence of two separate proton bumps in the cosmic-ray spectrum is a distinguishing feature of this model.
	
\begin{table}
    \centering
    \caption{Best-fit parameters for the single- and two-population source models with {\scshape EPOS-LHC} used as the hadronic-interaction model describing air-shower development. The $1\sigma$ uncertainties include the penalty factor for the total best-fit quality proposed in \cite{Rosenfeld:1975fy}. The ``1SC'' scenario is the benchmark model with only a single population of sources injecting mixed-composition cosmic rays. ``Population 1'' refers to the baseline source class that injects a mixed cosmic-ray flux of protons to iron, and ``Population 2'' denotes pure-proton sources. The best fit of UHECR spectrum and composition is given in the ``CR'' column, and the best fit after including neutrino and gamma-ray limits in the ``CR + MM'' columns. For the 2SC-uhecr model, the cosmic-ray best fit is compatible with existing multimessenger limits. Confidence intervals that extend to the edges of the sampled parameter range are indicated by an asterisk.}\label{tab:results_epos}
    \renewcommand{\arraystretch}{1.3}
    \begin{tabular}{l|rrrr}
        \hline
		Model	                &  1SC                      & \multicolumn{2}{c}{2SC-dip}                           & 2SC-uhecr   \\
			                    &  \                        & CR                        & CR + MM   & CR + MM  \\	\hline
		\textbf{Population 1}   & \                         & \                         & \                         & \  \\ \cline{1-1}
		$\rmax$ [EV]	        & $1.25^{+0.23}_{-0.19}$    & $1.5^{+0.5}_{-0.4}$       & $1.5^{+0.5}_{-0.4}$       & $1.5^{+0.5}_{-0.4}$    \\
		$\gamma$		        & $-2.5^{+1.0}_{-0}$        & $-1.20^{+0.22}_{-0.22}$   & $-1.41^{+0.44}_{-0.22}$   & $-1.41^{+0.22}_{-0.22}$   \\
		$m$		                & $1.9^{+0.6}_{-4.1}$       & $-2^{+2}_{-2}$            & $-1^{+1}_{-3}$            & $1^{+1}_{-2}$      \\
		L$_0$	[\tiny{$10^{44}\frac{\si{\erg}}{\si{\mega\parsec\cubed}~\si{\year}}$}\small]  & $5.6^{+1.0}_{-3.4}$	 & $2.0^{+1.0}_{-0.5}$     & $2.5^{+0.6}_{-1.0}$ & $3.77^{+0.06}_{-1.39}$  \\ \hline
        
 		$f_\text{p}^R [\%]$		& $6.6^{+15.4}_{-6.6}$      & $\approx0^{+6.5}_{-0}$    & $\approx0^{+8.6}_{-0}$    & $\approx0^{+7.2}_{-0}$  \\
 		$f_\text{He}^R [\%]$	& $48.1^{+7.6}_{-2.8}$      & $68.5^{+3.9}_{-7.3}$      & $66.9^{+4.6}_{-4.8}$      & $70.2^{+4.1}_{-4.4}$ \\
 		$f_\text{N}^R [\%]$	    & $40.1^{+4.9}_{-16.8}$     & $26.6^{+6.6}_{-3.5}$      & $28.0^{+4.5}_{-5.3}$      & $23.6^{+3.0}_{-4.8}$ \\
 		$f_\text{Si}^R [\%]$	& $4.8^{+0.7}_{-1.5}$       & $4.5^{+0.5}_{-0.5}$       & $4.8^{+0.3}_{-0.9}$       & $5.8^{+0.6}_{-1.5}$  \\
 		$f_\text{Fe}^R [\%]$	& $0.45^{+0.06}_{-0.17}$    & $0.38^{+0.10}_{-0.09}$    & $0.33^{+0.10}_{-0.05}$    & $0.42^{+0.07}_{-0.08}$  \\ \hline
        \textbf{Population 2}   & \                         & \                         & \                         & \  \\ \cline{1-1}
		$\emax^\ptwo$ [EeV]	    & \                         & $10^5$ (fix)              & $10^5$ (fix)              & $10$ (fix)  \\
		$\gamma^\ptwo$		    & \                         & $2.5^{+0.3}_{-0.3}$       & $2.5^{+0.3}_{-0.3}$       & $-0.25^{+0.50}_{-0.75}$   \\
		$m^\ptwo$		        & \                         & $6^{+0*}_{-10}$           & $4^{+1}_{-10*}$           & $-3^{+9*}_{-3*}$   \\
		L$_0^\ptwo$	[\tiny{$10^{44}\frac{\si{\erg}}{\si{\mega\parsec\cubed}~\si{\year}}$}\small]  & \		& $4.5^{+1.6}_{-4.0}$   & $1.8^{+0.6}_{-1.4}$   & $0.12^{+1.88}_{-0.06}$  \\ \hline
		$f^\ptwo(\SI{20}{\exa\electronvolt})\,[\%]$ & \             & $7.9^{+0}_{-0.9}$   & $7.0^{+0.8}_{-0.5}$   & $14.2^{+1.2}_{-0.5}$   \\
		$\chi^2 / \text{dof}$	& $101.0 / 29$              & $73.4 / 26$               & $74.4 / 26$               & $58.0 / 26$           \\
        \hline
    \end{tabular}
\end{table}

\section{Multimessenger Signal}\label{sec:results_multimessenger}
In the following, we discuss the predicted multimessenger signal produced through interactions with the CMB and the Extragalactic Background Light during the propagation of the cosmic rays. We focus on the 2SC-dip ``proton-dip'' model which predicts a large flux of cosmogenic neutrinos and UHE gamma rays. The multimessenger signal of the 2SC-uhecr model is briefly discussed at the end.

\subsection{2SC-dip}
\begin{figure}
    \centering
    \includegraphics[width=0.48\linewidth]{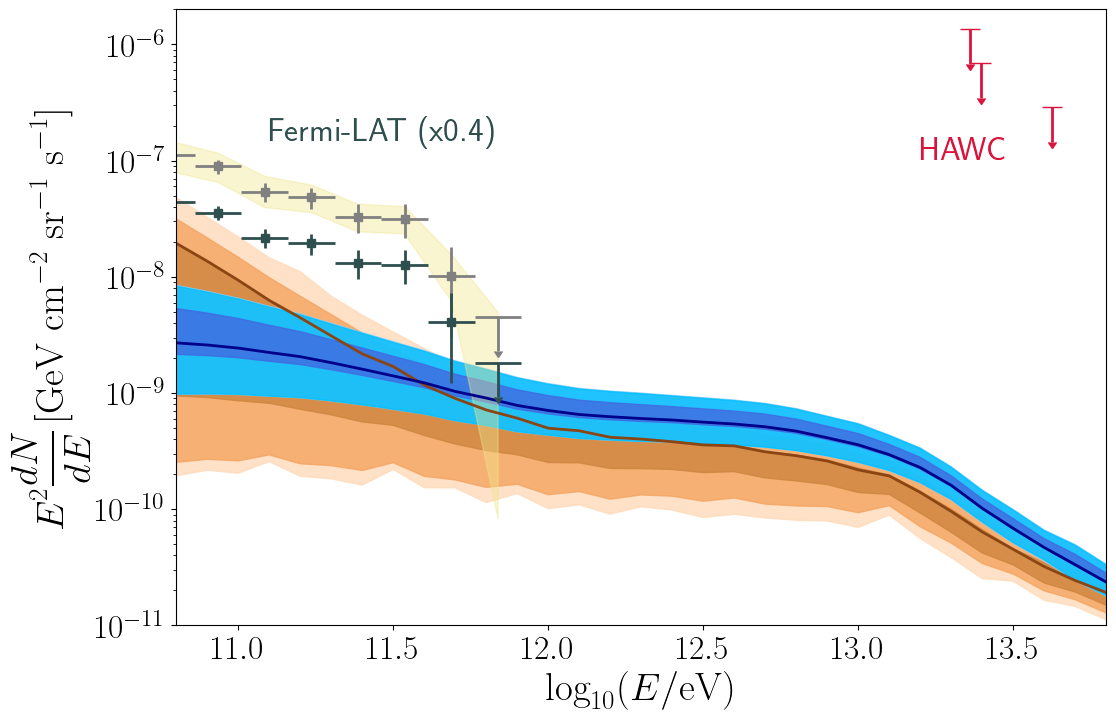}
    \includegraphics[width=0.48\linewidth]{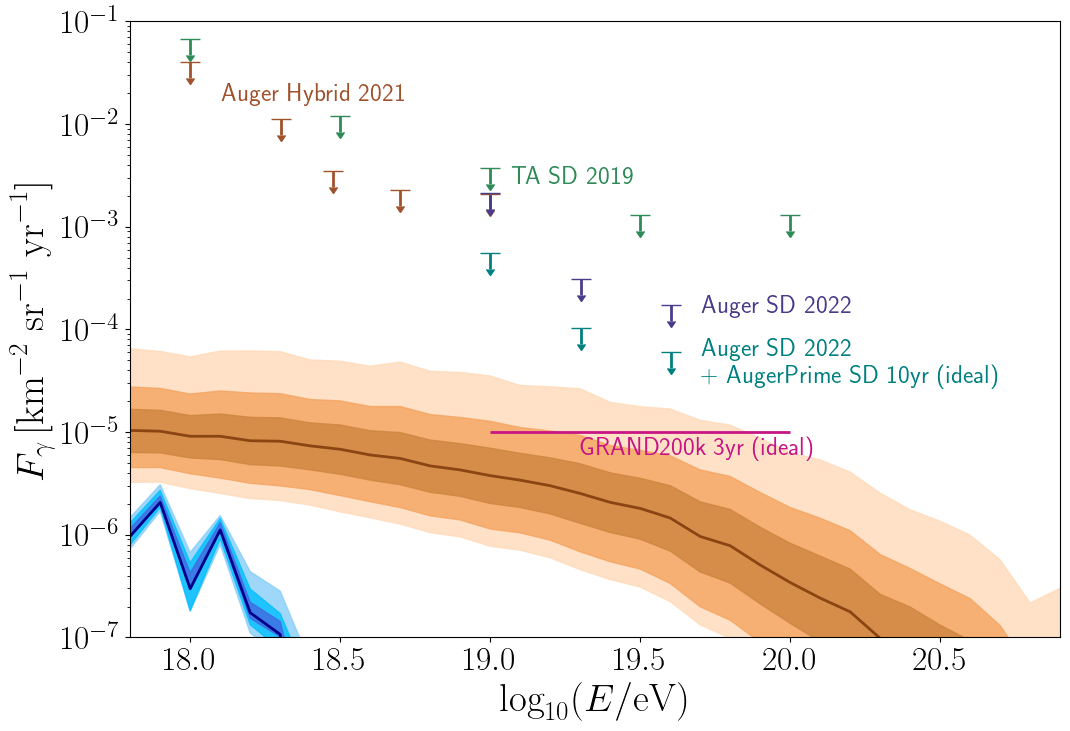}
    \caption{Predicted cosmogenic gamma-ray signal for the ``proton-dip'' model (2SC-dip), with {\scshape Epos-LHC} as hadronic interaction model, in the GeV-TeV (left) and EeV (right) energy range. The photon flux for each source class corresponding to the UHECR best fit (\cref{tab:results_epos}, 2nd column) is indicated by a solid line. The $1,2,3\sigma$ contours, under the condition that $\Delta\chi_\gamma^2<4$, are indicated by brown bands in decreasing intensity for the contribution from the additional protonic UHECRs, and by blue bands for the gamma-ray flux from the regular, mixed cosmic rays. These intervals do not include the best-fit penalty factor of \cite{Rosenfeld:1975fy}. Observations include the Fermi-LAT~\cite{Fermi-LAT:2014ryh} and HAWC~\cite{HAWC:2022uka} diffuse gamma-ray background in the GeV-PeV range, the 95\% upper limits at UHE of Auger~\cite{Savina:2021cva,PierreAuger:2022aty} and TA~\cite{TelescopeArray:2018rbt}, the optimistic 3-year sensitivity of the planned GRAND200k~\cite{GRAND:2018iaj}, and a combination of the latest Auger SD limit with the projected AugerPrime exposure for 10 years of observations under the assumption of 100\% photon selection efficiency and zero background.}\label{fig:EPOS_2SC_photons}
\end{figure}

Photons, electrons, and positrons produced with PeV-EeV energies in photohadronic interactions of the UHE protons interact with cosmic photon fields, leading to the development of electromagnetic cascades and reprocessing to lower energies. In the scenario of low-$\emax$ and mixed-composition cosmic-ray sources only, most of the gamma-ray signal is expected at GeV-PeV energies since the CR energies are insufficient for large interaction cross-sections with CMB photons. In this energy range (\cref{fig:EPOS_2SC_photons}, left), the predicted gamma-ray flux associated with the PP protons in our model is at a similar level to the flux expected from the mixed cosmic rays. Depending on the exact choice of source parameters, the combined gamma-ray flux of both populations can saturate the upper limit imposed by the re-scaled\footnote{Following~\cite{AlvesBatista:2018zui}, we re-scale the isotropic gamma-ray background reported by Fermi-LAT~\cite{Fermi-LAT:2014ryh} by a conservative factor of $\times0.4$ to account for the contribution from unresolved point sources which was estimated to be $\approx 68^{+9}_{-8}\%$ in~\cite{Lisanti:2016jub}.} Fermi-LAT flux at $\sim\SI{700}{\giga\electronvolt}$, however, the tension is not statistically significant. Most of the gamma-ray flux at $E_\gamma\gtrsim\,\SI{100}{\giga\electronvolt}$ is produced by the mixed-composition cosmic rays. At lower energies, the cosmogenic gamma rays are safely below the observed diffuse background flux.
    
The situation is more promising at ultra-high energies where the signal from the ordinary, mixed cosmic rays is expected to be very small. By construction, the protons injected at the PP sources have typical energies $E_\gamma>10^{18}\si{\electronvolt}$ and consequently large cross-sections for photo-pion production on the abundant CMB photons. The predicted UHE gamma-ray flux from the protons is therefore orders of magnitude above the flux produced by the mixed cosmic rays (\cref{fig:EPOS_2SC_photons}, right). It correlates inversely with the PP spectral index -- harder injection spectra result in more cosmogenic UHE photons. As indicated previously, hard injection spectra generally require strongly positive redshift evolutions to soften the observed spectrum. Present limits by Auger and TA are not constraining, even in the most optimistic scenario within $3\sigma$ uncertainties, however, the difference is not more than a factor of a few and it is clear that future detectors -- such as GRAND200k and AugerPrime -- will provide strong constraints for the viable PP spectral index and redshift evolution.

\begin{figure}
    \centering
    \includegraphics[width=0.48\linewidth]{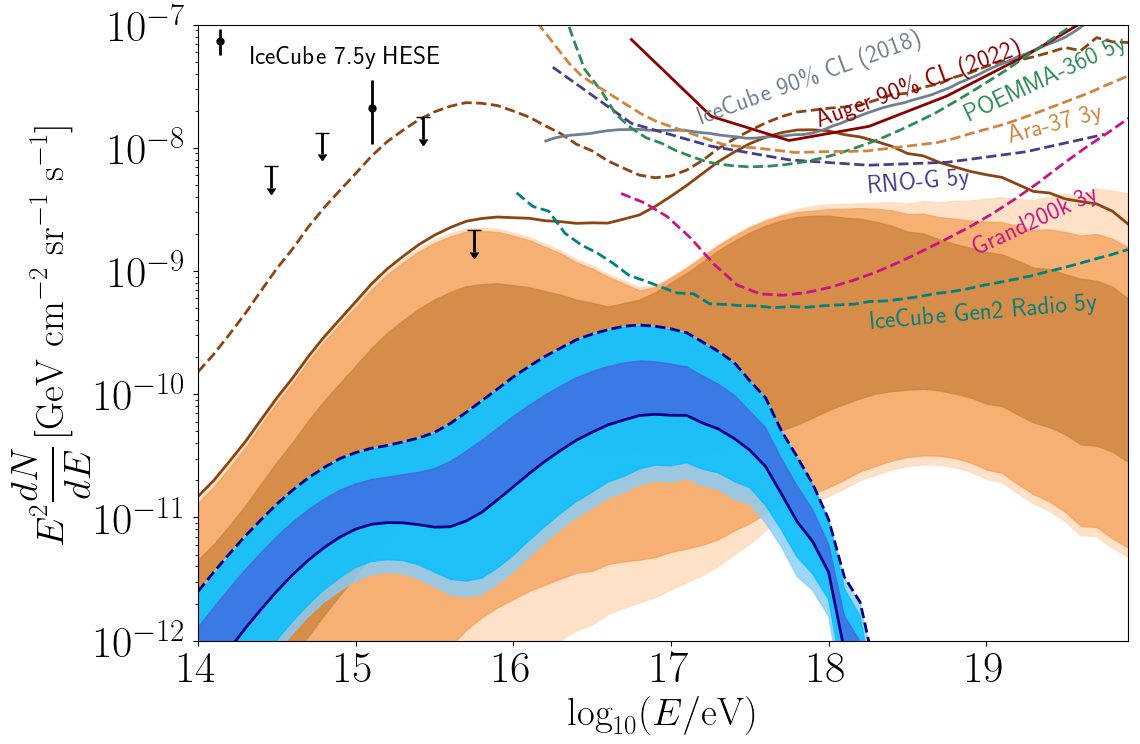}
    \includegraphics[width=0.48\linewidth]{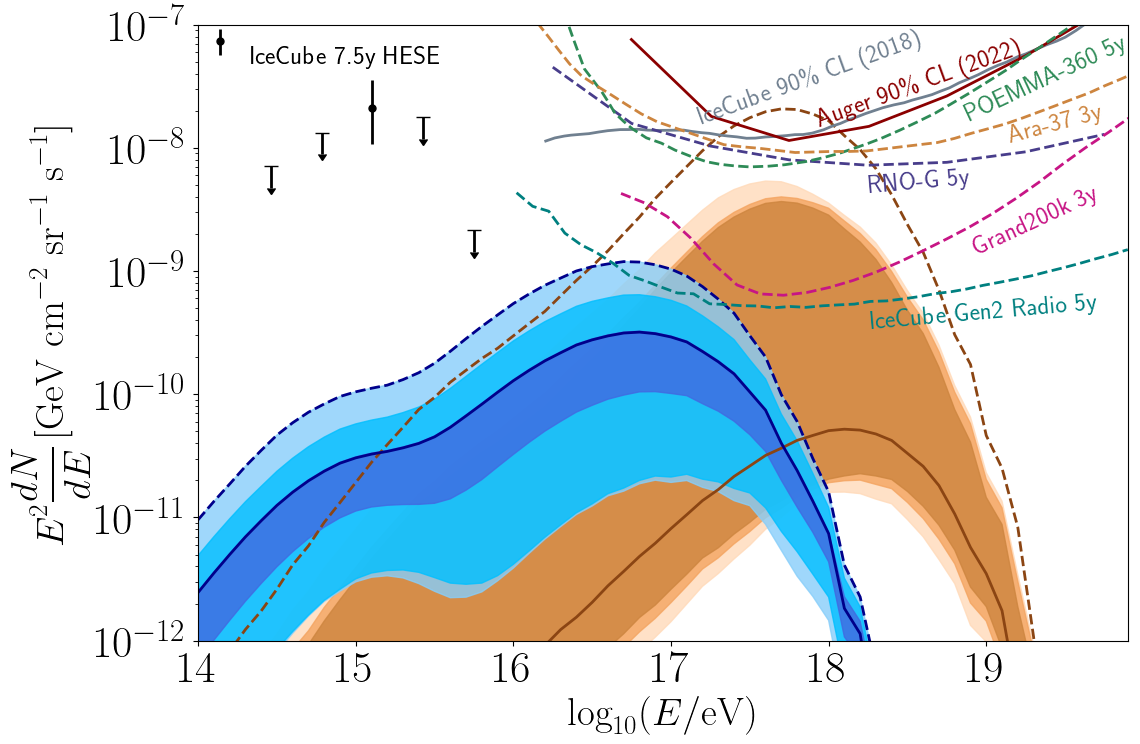}
    \caption{Same as \cref{fig:EPOS_2SC_photons} but for the predicted cosmogenic neutrinos in the 2SC-dip (left) and 2SC-uhecr model (right). The maximum allowed flux within $3\sigma$ of the best CR fit but without including the multimessenger penalty is shown as a dashed line of the respective colour. Solid lines indicate the neutrino flux corresponding to the cosmic-ray best fit without multimessenger constraints (\cref{tab:results_epos}, 2nd column). The shaded confidence intervals include the additional $\chi^2$ penalty from the existing neutrino limits. The IceCube HESE flux~\cite{IceCube:2020wum}, upper limits from IceCube~\cite{IceCube:2018fhm} and Auger~\cite{PierreAuger:2019ens,Pedreira:2021gcl}, and predicted sensitivities of planned detectors \cite{IceCube:2019pna,GRAND:2018iaj,ARA:2015wxq,Cummings:2020ycz} are shown as a reference.}\label{fig:EPOS_2SC_dip_neutrinos}
\end{figure}

The expected flux of cosmogenic neutrinos (\cref{fig:EPOS_2SC_dip_neutrinos}) is not well constrained by the cosmic-ray fit alone and can vary by approx.\ a factor of $1000$ within the $99.7\%$ confidence interval. In the most pessimistic case, when the redshift evolution of the proton sources is strongly negative, the neutrino flux produced by PP protons is subdominant to the neutrinos from the default CR population at all energies $E_\nu\lesssim\,\SI{1}{\exa\electronvolt}$ and the UHE flux is small. On the other hand, for strong redshift evolutions, the expected neutrino flux saturates the flux observed by IceCube in the few-PeV energy range and exceeds significantly the limits above \SI{10}{\peta\electronvolt} and at UHE. This includes the source configuration corresponding to the best UHECR spectrum and composition fit. By requiring that the neutrino limits are not violated ($\Delta\chi^2_\nu<4$) we can constrain the properties of the proton sources to
\begin{equation}\label{eq:results_constraints}
    \gamma^\ptwo \gtrsim 1.6 ~,~ m^\ptwo \lesssim 4 ~,{\rm and}~L_0^\ptwo \lesssim 10^{44.5}\,\frac{\si{\erg}}{\si{\mega\parsec\cubed\year}}\,.
\end{equation}
Irrespective of the total level, the predicted neutrino flux exhibits a characteristic double-bump profile, with the first peak at $E_\nu\approx\,\SI{5}{\peta\electronvolt}$ from photo-pion production of the cosmic-ray protons on the extragalactic background light, and the second peak at $E_\nu\approx\,\SI{1}{\exa\electronvolt}$ from photo-pion production on the less energetic, but more abundant, CMB photons. Due to the soft spectrum of the UHE protons, both peaks are present at the same time and the UHE neutrino limits can be used to constrain the contribution of this cosmogenic neutrino flux to the observed IceCube HESE flux at \SI{1.3}{\peta\electronvolt} to $f_\text{HESE}^\ptwo\lesssim20\%$.

\subsection{2SC-uhecr}
In the ``UHECR best fit'' model, the maximum proton energy is below the required level for photo-pion production with the bulk of CMB photons and the expected multimessenger signal is low. UHE gamma rays are at least three orders of magnitude below existing limits and at GeV-TeV energies, the contribution is subdominant compared to the cosmogenic photons from the MIX cosmic rays. The total contribution to the Fermi-LAT IGRB is $<50\%$ even in the most optimistic scenario, although the upper limit in the highest energy bin is approximately saturated.
    
While the neutrino signal of the UHE protons at the best fit is subdominant to the neutrinos from the mixed-composition cosmic rays, the shape of the neutrino spectrum is of particular interest. Unlike for the 2SC-dip model, few protons are present at lower energies and the low-energy peak originating from interactions with EBL photons is therefore absent. Only the peak from photo-pion production on the CMB remains. In this scenario, the observed IceCube neutrino flux at PeV energies and below, and the possible UHE neutrino flux are decoupled. It is possible, for strongly positive redshift evolutions of the proton sources, to produce a large neutrino flux at UHE with a negligible contribution to the IceCube HESE flux. Redshift evolutions stronger than $m^\ptwo\approx4$ can be excluded by the current UHE neutrino limits of IceCube and Auger.

\section{Exotic Flux Recovery Scenario (2SC-rec)}\label{sec:results_recovery}
\begin{figure}
    \centering
    \includegraphics[width=0.48\linewidth]{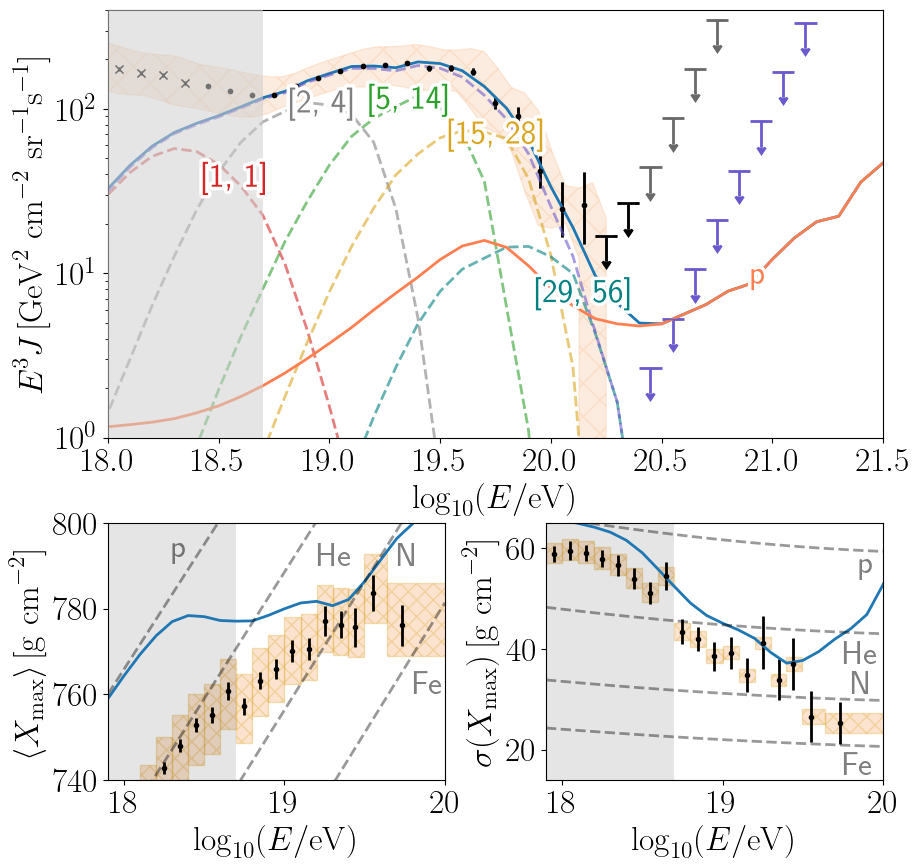}
    \caption{Same as \cref{fig:EPOS_2SC_CR} but for the ``flux recovery'' 2SC-rec scenario. Auger 90\% upper limits above $10^{20.4}\,\si{\electronvolt}$ were derived assuming an energy-independent exposure of \SI{60400}{\kilo\meter\squared\year\steradian}~\cite{PierreAuger:2020qqz}. Expected 90\% upper limits for GCOS (40k) after 10 years of operation ($\epsilon\sim10^6\,\si{\kilo\meter\squared\year\steradian}$~\cite{Coleman:2022abf}) are shown in purple.}\label{fig:EPOS_2SC_CR_recovery}
\end{figure}
A combination of the 2SC-dip and 2SC-uhecr models is provided by a proton source population with large maximum energy, $\emax^\ptwo=10^5\,\si{\exa\electronvolt}$, as in the ``proton-dip'' model, and hard injection spectrum, $\gamma^\ptwo=1$, similar to the ``UHECR'' model. The quality of the UHECR fit (\cref{tab:results_epos_recovery}) is reduced compared to the other two models and approaches the baseline single-source-class model. Compared to the dip model, the potential for fit improvement is limited since the protons contribute only at the highest energies, while the position of the observed proton peak is at too-high energies to provide an improvement of similar magnitude as in the 2SC-uhecr model. However, an interesting feature in the form of a ``flux recovery'' at trans-GZK energies can be observed (\cref{fig:EPOS_2SC_CR_recovery}). We refer to this third model as the ``recovery'' or 2SC-rec model.

A recovery is only possible if the nearest source(s) is(are) located within the GZK volume at no more than $\sim\,\SI{20}{\mega\parsec}$~\cite[e.g.][]{Gaisser:2016uoy} as otherwise, the GZK cutoff provides a natural suppression of the observable flux above $\sim10^{19.7}\,\si{\electronvolt}$. Such a spectral recovery is not necessarily connected to a large UHE neutrino signal. In addition to a high $\emax^\ptwo$ and hard proton source spectra, the latter also requires strong redshift evolution of the source emissivity which is not a pre-requisite for a CR flux recovery. However, the observation of neutrinos with energy above $10^{19}\si{\electronvolt}$ by future extremely-UHE neutrino detectors such as PUEO~\cite{PUEO:2020bnn} would provide a strong hint for the existence of a sizeable UHECR flux recovery beyond the GZK cutoff.

We have noted in \cref{sec:results_multimessenger} for large $\emax^\ptwo$ that hard proton source spectra are excluded by existing neutrino limits, under the condition that only source configurations within $3\sigma$ of the best fit to the UHECR spectrum and composition under the 2SC-dip model are considered. If this limitation is lifted, such as for the 2SC-rec model, we can identify scenarios where the predicted neutrino flux is sufficiently below existing limits, i.e. $\Delta\chi^2_\nu<4$. This constrains the redshift evolution of the proton sources to $m^\ptwo\lesssim3$ ($\lesssim2$ if gamma-ray limits are included).

In contrast to the 2SC-dip model, the combination of hard injection spectrum, high maximum proton energy and uniform source distribution with minimum distance $z_\text{min}=10^{-3}$, results in an increased flux of cosmogenic UHE gamma rays. We find that for spectral indices harder than $\gamma^\ptwo\lesssim1$ all possible realisations of the source model are excluded by the existing Auger UHE photon limits~\cite{Savina:2021cva,PierreAuger:2022aty}. We conclude that the joint consideration of neutrino and UHE gamma-ray limits severely constrains the allowed proton injection spectrum and, by extension, the maximum allowed flux recovery from this second source population above the GZK cutoff. This motivates our choice of $\gamma^\ptwo=1$ as benchmark spectral index for the 2SC-rec model. The spectrum and composition corresponding to the maximum recovery allowed by the cosmic-ray fit and multimessenger constraints are shown in \cref{fig:EPOS_2SC_CR_recovery}. The preferred source parameters stay unchanged except for the PP redshift evolution and luminosity density. Finally, we comment that the projected sensitivity of the proposed \emph{Global Cosmic Ray Observatory} (GCOS)~\cite{Coleman:2022abf} would place strong constraints on the allowed UHE flux recovery.

Our results for this scenario can look similar to what is expected from propagation models involving Lorentz invariance violation (LIV) (compare with \cite{Aloisio:2000cm,Scully:2008jp,PierreAuger:2021tog}, for example), which can suppress the photopion interaction at high energies and strengthen the flux recovery. If a recovery is observed, it is therefore prudent to investigate whether this is due to LIV or a 2SC-dip/recovery scenario. Clear differences between LIV models and a 2SC-dip/recovery scenario are the expected arrival directions as well as the expected cosmogenic neutrino and photon fluxes.

\begin{table}
\centering
    \caption{Same as \cref{tab:results_epos} but for the extreme 2SC-recovery model. The best-fit parameters of the mixed-composition sources are given for the CR-only fit, but the preferred values for the CR+MM scenario are compatible within quoted uncertainties. }\label{tab:results_epos_recovery}
    \renewcommand{\arraystretch}{1.3}
    \begin{tabular}{l|rr|rr}
        \hline
		\textbf{Population}     & MIX                       & $f_A^R [\%]$          & \multicolumn{2}{c}{Pure-Proton}   \\
		                        &                           &                       & CR                & CR + MM       \\
		\hline
		$\rmax$ [EV]	        & $1.4^{+1.0}_{-0.6}$       & $10.5^{+7.6}_{-10.2}$ & $10^5$ (fix)      & $10^5$ (fix)  \\
		$\gamma$		        & $-1.7^{+0.4}_{-0.4}$      & $55.7^{+4.5}_{-3.6}$  & $1$ (fix)         & $1$ (fix)     \\
		$m$		                & $0^{+1}_{-1}$             & $29.3^{+4.8}_{-3.4}$  & $6^{+0*}_{-2}$    & $2^{+1}_{-4}$ \\
		L$_0$	[\tiny{$10^{44}\frac{\si{\erg}}{\si{\mega\parsec\cubed}~\si{\year}}$}\small]  & $3.3^{+0.8}_{-0.6}$	 & $4.3^{+1.0}_{-0.7}$ & $12.0^{+1.7}_{-8.1}$  & $1.5^{+0.6}_{-1.1}$    \\ 
		                        &                           & $0.22^{+0.01}_{-0.02}$ &                  &               \\
		\hline
		$f^\ptwo(\SI{20}{\exa\electronvolt})\,[\%]$ & \     &                       & $2.6^{+0}_{-0}$   & $1.46^{+0.06}_{-0.20}$ \\
		$\chi^2 / \text{dof}$	& \                         &                       & $87.3 / 27$       & $91.9 / 27$   \\
        \hline
    \end{tabular}
\end{table}

\section{Discussion and Comparison to Previous Studies}\label{sec:paper_comparison}
Here we discuss the implications of fitted parameters and compare our findings to past works investigating a subdominant proton flux at the highest energies.

A similar conclusion in terms of the allowed UHE proton fraction for {\scshape Epos-LHC} versus {\scshape Sibyll2.3c} was reached in the recent paper by Muzio et al.\ \cite{Muzio:2023skc}. The best fit obtained in this study (not explicitly discussed in their paper) is qualitatively similar to our 2SC-uhecr model with hard proton injection spectrum and low maximum energies\footnote{private communication with M.\ Stein Muzio} (see also \cite{Muzio:2019leu}). Their reported, best-fit observed proton fraction of the integral cosmic-ray flux above \SI{30}{\exa\electronvolt} for typical astrophysical source evolutions is $5-10\%$ and $2-3\%$ for {\scshape Epos-LHC} and {\scshape Sibyll2.3c} respectively. These values are compatible with our preferred integral fractions, $F^\text{p}(\geq\SI{30}{\exa\electronvolt})=10.2^{+1.4}_{-1.5}\%\,/\, 3.1^{+2.6}_{-0.9}\%$. A direct comparison is difficult, however, since the authors assumed a mono-elemental injection of Silicon-like nuclei at the MIX sources and also included in-source photohadronic interactions, and the predicted cosmic-ray flux at Earth is not provided.

A solution to the two-population model with source parameters similar to our 2SC-dip best fit was found by Das et al.\ \cite{Das:2020nvx} who report a best-fit proton fraction of the observed flux in the highest energy bin of approx.\ $1-3\%$. This corresponds to $20-25\%$ at $E_\text{ref}=\,\SI{20}{\exa\electronvolt}$ -- a contribution that we found to be in strong tension with the observed UHECR composition, in particular the variance of shower maxima. They assumed a maximum source distance of $z_\text{max}=1$ which results in a conservative estimate of the associated flux of cosmogenic neutrinos. We consider, instead, sources out to redshift $z_\text{max}=4$, based on the approximate redshift evolution of source emissivities for probable astrophysical sources of UHECRs, resulting in a neutrino flux exceeding their prediction by more than an order of magnitude at $E_\nu=\SI{1}{\exa\electronvolt}$. This enables us to use existing upper limits on the UHE neutrino flux by Auger and IceCube to significantly constrain the redshift evolution of the PP source population emissivity. Unlike the present work, the expected flux of cosmogenic gamma-rays is not discussed extensively in \cite{Das:2020nvx}. The authors do not recover the best fit of the UHECR spectrum and composition that we identify in our 2SC-uhecr model since they only consider soft spectral indices of the proton sources, $\gamma^\ptwo\geq2.2$, and no mention is made of a possible flux recovery beyond the GZK cutoff.

A recent study by the \textsc{Auger Collaboration}~\cite{PierreAuger:2022atd} also investigated the co-existence of several source populations to explain the entire UHECR spectrum and composition above and below the ankle simultaneously. While the focus of that paper is somewhat different, their \textit{scenario 1} resembles our proton-bump model and the best-fit parameters are generally compatible. Mild disagreement can be identified for the preferred spectral index of the proton sources, which they predict to be much softer, and the redshift evolution of the mixed-composition sources, which they predict to be substantially stronger. Both of these are likely related the lower limit on the energy range in~\cite{PierreAuger:2022atd}. They require a larger proton flux below the ankle to explain the entire observed flux while we only use those data points as upper limits. However, their result depends on the assumptions for the sub-ankle nitrogen flux component which they have fixed \textit{ad-hoc}.

Important information about the potential sources of the UHE pure-proton flux can be gained from the total emissivity $L_0^\text{PP}$ (luminosity density) required by the UHECR fit. Although the cosmic-ray emissivity of astrophysical objects is generally not known, other observable properties such as gamma-ray and X-ray emissivities can be used for relative calibration. For a summary of population emissivities see \cite{Murase:2018utn}. Assuming equipartition of the available energy budget into gamma rays / X-rays and cosmic rays, we observe that all typically considered source classes (gamma ray bursts, tidal disruption events, starburst galaxies, active galactic nuclei, BL Lacertae, flat-spectrum radio quasars, and radio galaxies) can satisfy the emissivity required of the pure-proton sources in the 2SC-uhecr and 2SC-dip models, although gamma-ray bursts and tidal disruption events are marginally challenged in the latter scenario. For the extreme 2SC-recovery model, only the entire AGN population and the population of all BL Lacs can easily meet the required emissivity. GRBs and TDEs, in contrast, are excluded unless their cosmic-ray emissivity exceeds the observed gamma-ray emissivity by at least a factor of ten. FSRQs and radio galaxies sit close to the minimum luminosity density required by the cosmic-ray fit.

Given the hard spectrum and high maximum energy, it might be challenging that the UHE proton flux predicted by the 2SC-rec model is produced by astrophysical accelerators. An alternative explanation for the spectrum could be provided by the decay of hypothetical super-heavy dark matter (SHDM) with masses up to the Planck mass~\cite{Berezinsky:1997hy,Kuzmin:1997jua,Sigl:1998vz,Bhattacharjee:1999mup,Ellis:2005jc,Kalashev:2008dh,Aloisio:2015lva,Supanitsky:2019ayx}. These heavy particles can be produced gravitationally during the early stages of the Universe, e.g.\ as part of the reheating epoch from a hypothesised, decaying inflaton field, or from coherent oscillations of this field before the inflation phase~\cite{Kofman:1994rk,Felder:1998vq,Chung:1998rq}. If they never reached thermal equilibrium after production and the lifetime is larger than the age of the Universe then these heavy relics can provide a possible explanation for observed DM densities~\cite{Aloisio:2015lva}. Similar to the original proton-dip model~\cite{Berezinsky:1988wi}, ``top-down'' scenarios of decaying SHDM are disfavoured as the single origin of the observed UHECR flux~\cite{PierreAuger:2016kuz,Rautenberg:2021vvt}, and it was shown that decaying SHDM cannot explain the detected high-energy IceCube neutrino events if a hadronic decay channel is considered~\cite{Kuznetsov:2016fjt, Cohen:2016uyg,Kachelriess:2018rty}. Still, a subdominant contribution to the observed UHECR flux, and a possible flux recovery due to very hard decay spectra are not fully excluded. Crucially, existing upper limits on the post-GZK cosmic-ray flux provide only weak constraints on the allowed flux recovery, and UHE photon limits prove superior for $M_\text{DM} < 10^{14}\,\si{\giga\electronvolt}$~\cite{Supanitsky:2019ayx}.

We do not investigate the SHDM scenario further, however, we wish to point out several key differences compared to our assumed source model. If the additional protons are produced in the decay of super-heavy dark matter, a substantial anisotropy in arrival directions and extremely local production of the observed UHECRs should be expected since the signal is predicted to be dominated by dark matter in the Milky Way with a particular clustering around the Galactic centre~\cite{PierreAuger:2022ubv}. This is in sharp contrast to our proposed continuous distribution of sources in redshift up to $z_\text{max}=4$ and minimum source distance of $\sim\SI{4}{\mega\parsec}$. Consequently, in the SHDM scenario, the expected flux of cosmogenic neutrinos and low-energy gamma rays is severely reduced. In addition, we only consider the cosmogenic production of neutrinos and gamma rays while in the SHDM model the multimessenger signal is likely dominated by production during the decay of the dark matter.

\section{Summary and Conclusions}\label{sec:conclusions}
In this work, we have investigated the possible existence, and allowed parameter space, for an additional, proton-dominated component of UHECRs, produced by an independent astrophysical source population. We have presented the maximum contribution of such a population to the UHECR flux at Earth, taking into account the fit to the UHECR spectrum and composition-sensitive observables. In addition, we have derived predictions for the spectral shape and redshift evolution of the independent UHE-proton population model as well as the expected secondary neutrino and photon fluxes produced by UHECR interactions and their detectability.

This analysis was performed for two distinct choices of the maximum proton energy. For sources with maximum energy far beyond the GZK limit (2SC-dip model), the proton spectrum at Earth reproduces the predictions of the classic ``proton-dip'' model~\cite{Berezinsky:1988wi}, albeit with the proton flux subdominant to the contribution of the principal, mixed-composition cosmic rays. If instead maximum energies below $10^{19.7}\,\si{\electronvolt}$ are assumed (2SC-uhecr model), the cosmic-ray fit is improved by $\Delta\chi^2\approx-15$ but the source spectrum must be hard and the associated multimessenger signature is generally small. In both scenarios, the redshift evolution of the proton sources cannot be constrained by the cosmic-ray fit alone.

We find that the maximum proton contribution to the observed, diffuse UHECR flux depends strongly on the choice of hadronic interaction model for the interpretation of the extensive air showers, and on the maximum proton energy. With {\scshape Sibyll2.3c} a proton fraction of $\lesssim 1\%$ is expected at $\SI{20}{\exa\electronvolt}$ in the 2SC-dip model and the improvement over the baseline model is negligible. Under the 2SC-uhecr model, a contribution of $2-5\%$ is predicted with a minor $1.1\sigma$ significance compared to the baseline one-population model. Assuming {\scshape Epos-LHC} instead, for the 2SC-dip model, approximately $8\%$ of the UHECR flux is expected to be protons, with the contribution nearly constant over the entire energy range above the ankle. For the 2SC-uhecr model, where $\emax^\text{p}=\SI{10}{\exa\volt}$, the contribution to the observed UHECR flux peaks around $E_\text{ref}\approx\SI{20}{\exa\electronvolt}$ at up to $15\%$, but the relative proton fraction decreases rapidly for energies away from the peak and the source spectra are required to be hard. The improvement of the two-population model over the baseline single-population scenario is $2.2\sigma$ (2SC-dip) and $3.7\sigma$ (2SC-uhecr).

We demonstrated that for our fiducial high-$\emax^\text{p}$ model a distinguishing feature of the independent UHE proton component is a soft spectral index ($\gamma=2.5\pm0.3$), which can be tested by AugerPrime or other facilities with event-by-event mass determination capabilities. In addition, the cosmogenic neutrino and UHE photon fluxes produced by this component are substantial and dominate over those from the mixed-composition population. Current neutrino upper limits from IceCube and Auger already weakly constrain the available parameter space for the proton population from the fit to the UHECR data alone.

Finally, as an ``exotic'' scenario, we have considered proton sources with high maximum energy $\emax^\text{p}\gg10^{20}\, \si{\electronvolt}$ and hard spectral index. We find that existing limits on the neutrino and UHE gamma-ray flux constrain the proton spectral index to $\gamma^\text{p}\gtrsim1$ and therefore provide an upper limit on the possible cosmic ray flux beyond the GZK cutoff. However, a significant recovery is still allowed.


\acknowledgments
We thank Bj\"orn Eichmann, Michael Kachelriess, Marco Muzio, Pavlo Plotko, and Michael Unger for useful discussions. AvV acknowledges support from Khalifa University's FSU-2022-025 grant.

\section*{Data Availability}
No new observational data was generated as part of this study. The cosmic-ray, gamma-ray and neutrino propagation was simulated with the \textsc{CRPropa\,3} software package~\cite{Batista:2016yrx,AlvesBatista:2022vem}, which is publicly available from \href{https://crpropa.desy.de}{crpropa.desy.de}.

\appendix
\section{Two-Population Fit with {\scshape Sibyll2.3c}}\label{apx:sibyll_fit}
\begin{figure}
    \centering
    \includegraphics[width=0.48\linewidth]{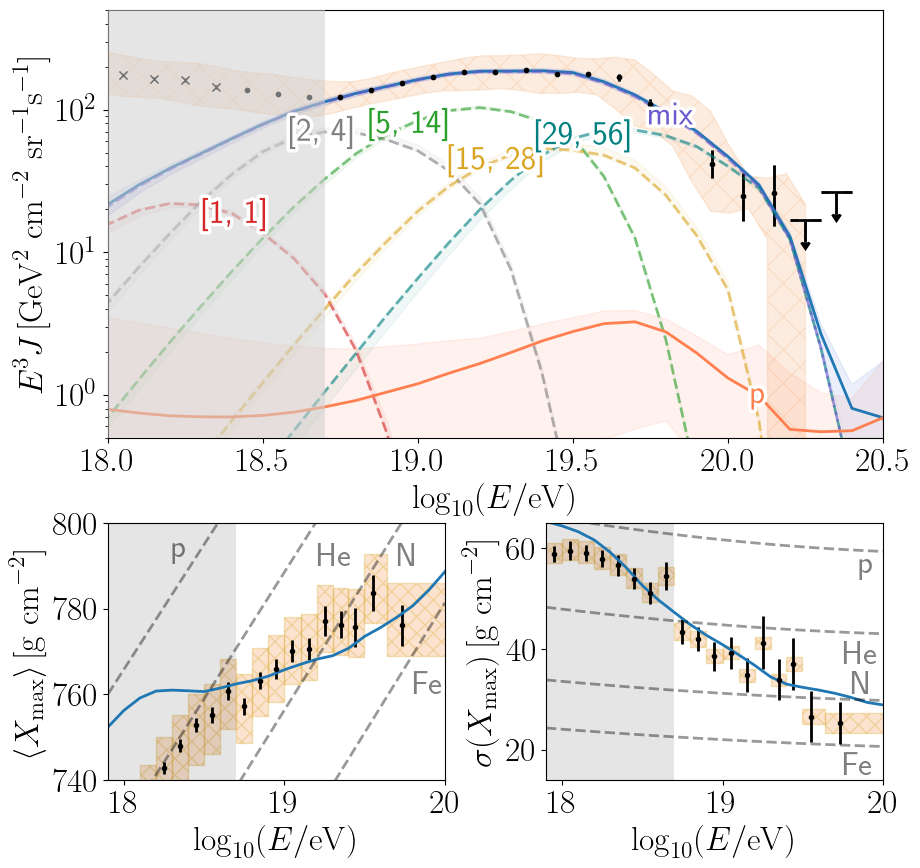}
    \includegraphics[width=0.48\linewidth]{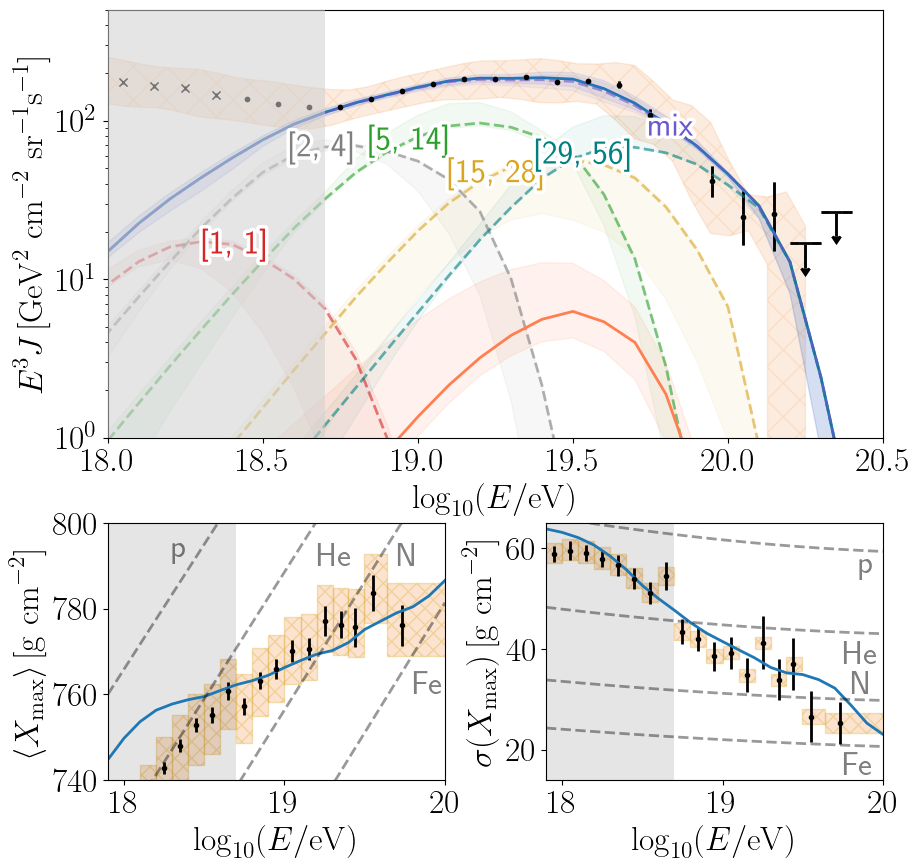}
    \caption{Same as \cref{fig:EPOS_2SC_CR} but with {\scshape Sibyll2.3c} as hadronic interaction model. Left: ``proton-dip'' (2SC-dip). Right: ``UHECR'' best fit (2SC-uhecr). The best-fit parameters are listed in~\cref{tab:results_sibyll}.}\label{fig:Sibyll_2SC_CR}
\end{figure}

We find that the level of additional protons at UHE compatible with observations critically depends on the considered hadronic interaction model. With {\scshape Sibyll2.3c}, the maximum improvement in fit quality over the single-population model is marginal for the high-$\emax^\ptwo$ 2SC-dip model, and reaches a significance of $1.1\sigma$ for the 2SC-uhecr scenario. Best-fit PP contributions to the observed flux at \SI{20}{\exa\electronvolt} are at a level of $1.1^{+0.1}_{-1.0}\%$ and $2.4^{+3.1}_{-0.2}\%$ respectively (\cref{tab:results_sibyll}). The cosmic-ray best fit for both scenarios is shown in \cref{fig:Sibyll_2SC_CR}.

\begin{figure}
    \centering
    \includegraphics[width=0.48\linewidth]{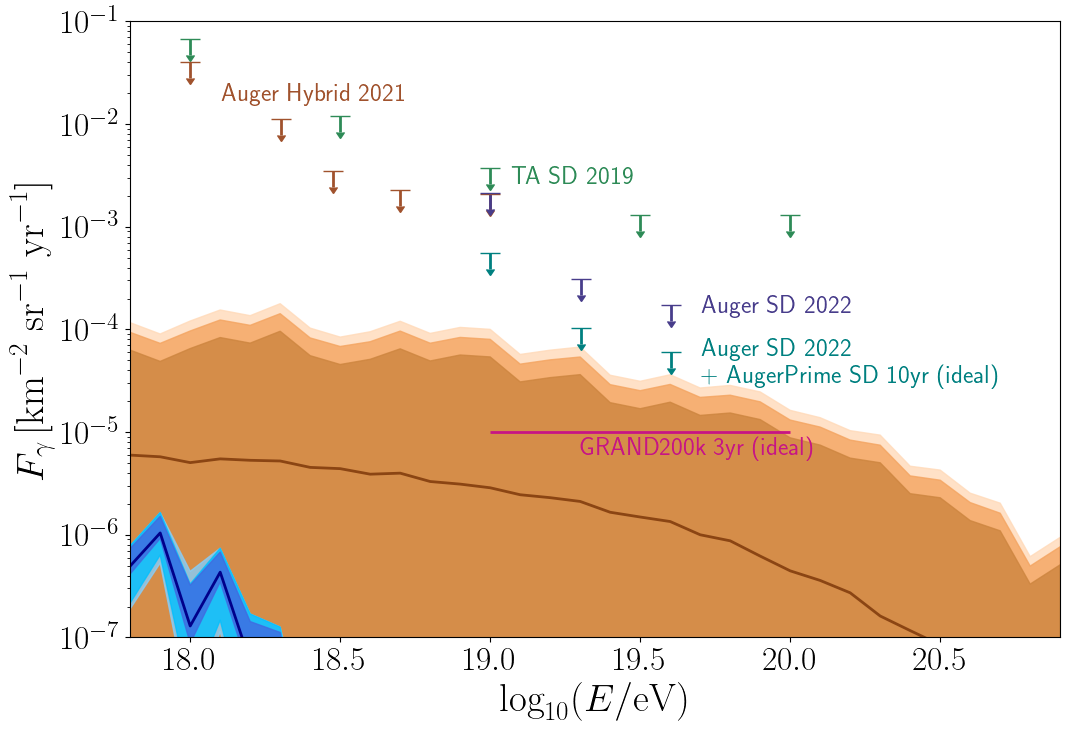}
    \includegraphics[width=0.48\linewidth]{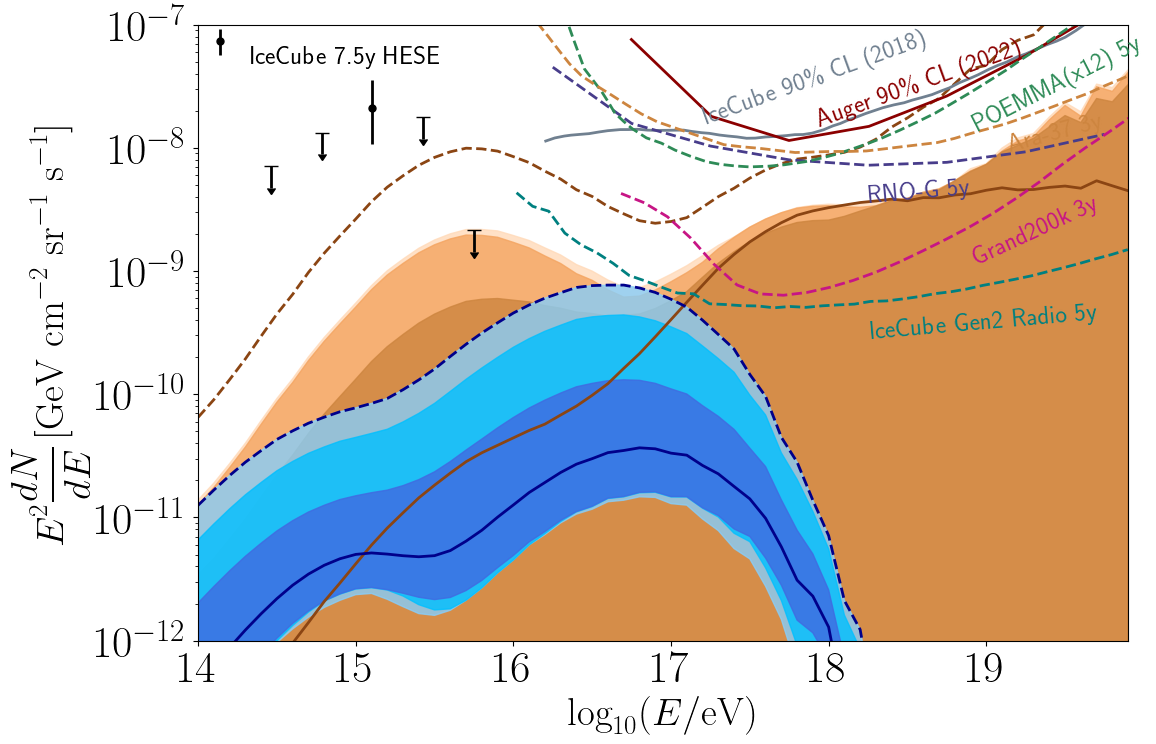}
    \caption{Same as \cref{fig:EPOS_2SC_photons,fig:EPOS_2SC_dip_neutrinos} for the ``proton-dip'' model (2SC-dip) but with {\scshape Sibyll2.3c} as hadronic interaction model. Left: UHE gamma rays. Right: neutrinos. The jagged upper limit of the UHE gamma-ray flux is a result of limited statistics in the numerical simulation.}\label{fig:apx_sibyll_mm}
\end{figure}

The pronounced differences between the two models are related to the quality of the single-population fit, in particular the $\mxmax$ fit. While with {\scshape Sibyll2.3c} a good fit of all observables can be achieved, {\scshape Epos-LHC} is not able to provide a good fit of the spectrum and $\mxmax$ at the same time. Consequently, for the latter, there is more room for improvement of the fit by an additional CR component, whereas for the two-population model with {\scshape Sibyll2.3c} the potential is limited. Although the contribution to the observed UHECR flux is generally low with {\scshape Sibyll2.3c}, in the 2SC-dip model the associated flux of cosmogenic neutrinos and UHE gamma rays can still be dominant compared to the fluxes derived from the mixed-composition cosmic rays (\cref{fig:apx_sibyll_mm}). However, they do not constrain the allowed PP source parameters except for $L_0^\ptwo\lesssim10^{44.3}\,\si{\erg\per\mega\parsec\cubed\per\year}$.

\begin{table}
\centering
    \caption{\label{tab:results_sibyll}
    Same as \cref{tab:results_epos} but for {\scshape Sibyll2.3c} as hadronic interaction model. The 2SC-dip best fit is excluded by the neutrino limits at $\Delta\chi^2_\nu\approx4$, however, compatibility is obtained for $m^\ptwo=6\to5$. For the 2SC-uhecr scenario, the CR best fit is again compatible with the multimessenger constraints.}
    \renewcommand{\arraystretch}{1.3}
    \begin{tabular}{l|rrr}
    \hline
	Model	                & 1SC                       & 2SC-dip                   & 2SC-uhecr             \\
	\hline
	\textbf{Population 1}   &                           &                           &  \\ \cline{1-1}
	$\rmax$ [EV]	        & $1.7^{+0.9}_{-0.6}$       & $1.7^{+0.9}_{-0.6}$       & $2.0^{+0.7}_{-0.9}$   \\
	$\gamma$		        & $0.15^{+0.59}_{-0.29}$    & $0.15^{+0.59}_{-0.29}$    & $0.54^{+0.22}_{-0.87}$ \\
	$m$		                & $-4^{+4}_{-1}$            & $-3^{+3}_{-3*}$           & $-6^{+2}_{-0*}$       \\
	L$_0$	[\tiny{$10^{44}\frac{\si{\erg}}{\si{\mega\parsec\cubed}~\si{\year}}$}\small]  &	$1.61^{+1.12}_{-0.12}$	&	$1.74^{+1.0}_{-0.4}$ &	$1.41^{+0.19}_{-0.12}$  \\ \hline
			
 	$f_\text{p}^R [\%]$		& $\approx0^{+0}_{-0}$      & $\approx0^{+0}_{-0}$      & $\approx0^{+0}_{-0}$  \\
 	$f_\text{He}^R [\%]$	& $66.9^{+2.5}_{-13.2}$     & $63.4^{+6.1}_{-10.9}$     & $70.4^{+0.7}_{-7.2}$  \\
 	$f_\text{N}^R [\%]$		& $28.5^{+11.3}_{-2.3}$     & $31.8^{+9.3}_{-5.6}$      & $25.3^{+8.7}_{-0.8}$  \\
 	$f_\text{Si}^R [\%]$	& $3.5^{+1.5}_{-0.1}$       & $3.5^{+1.4}_{-0.4}$       & $3.2^{+0.2}_{-2.1}$   \\
 	$f_\text{Fe}^R [\%]$	& $1.1^{+0.4}_{-0.1}$       & $1.3^{+0.3}_{-0.3}$       & $1.1^{+0.7}_{-0}$     \\ \hline
    \textbf{Population 2}   &                           &                           &                       \\ \cline{1-1}
	$\emax^\ptwo$ [EeV]	    &                           & $10^5$ (fix)              & $10$ (fix)            \\
	$\gamma^\ptwo$		    &                           & $1.9^{+1.1*}_{-3.4*}$     & $-1.3^{+1.3}_{-0.3*}$ \\
	$m^\ptwo$		        &                           & $6.0^{+0*}_{-12.0*}$      & $-6^{+12*}_{-0*}$     \\
	L$_0^\ptwo$	[\tiny{$10^{44}\frac{\si{\erg}}{\si{\mega\parsec\cubed}~\si{\year}}$}\small]  &		&	$0.7^{+5.9}_{-0.7}$     &	$0.012^{+0.401}_{-0.001}$  \\ \hline
	$f^\ptwo(\SI{20}{\exa\electronvolt})\,[\%]$ &       & $1.1^{+0.1}_{-1.0}$       & $2.4^{+3.1}_{-0.2}$   \\
	$\chi^2 / \text{dof}$	& $59.7 / 29$               & $58.3 / 26$               & $52.2 / 26$           \\
        \hline
    \end{tabular}
\end{table}

\section{Proton Maximum Energy}\label{apx:proton_emax}
\begin{figure}
    \centering
    \includegraphics[width=\linewidth]{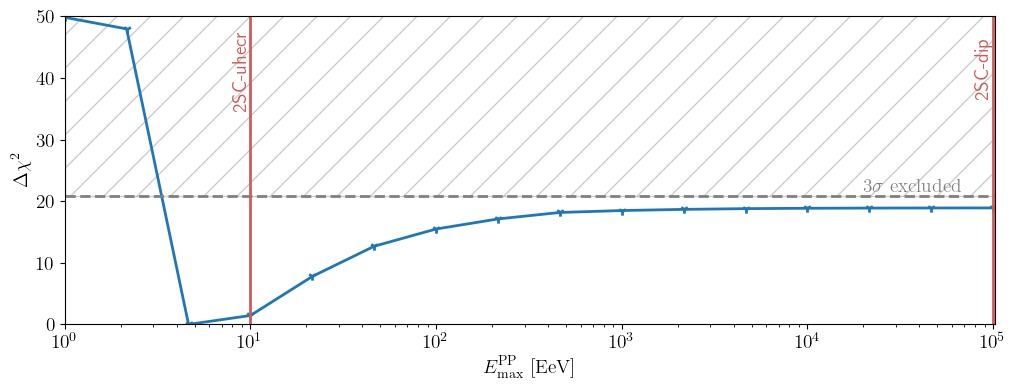}
    \caption{Fit quality ($\Delta\chi^2$) compared to the best fit as function of the proton maximum energy $\emax^\ptwo$. Values on the order of $\SI{10}{\exa\electronvolt}$ are preferred (corresponding to the 2SC-uhecr model); however, trans-GZK maximum energies cannot be rejected at appreciable significance (including the penalty for the quality of the global best fit \cite{Rosenfeld:1975fy}).}\label{fig:apx_proton_emax}
\end{figure}

Adding another free parameter in the form of the maximum proton energy $\emax^\ptwo$ to our fit is computationally prohibitive for our regular parameter pixelisation used in \cref{sec:results_UHECR}. However, if the resolution is reduced then a scan over $\emax^\ptwo$ is possible. The results are shown in \cref{fig:apx_proton_emax}. We note that $\emax^\ptwo\lesssim\SI{3}{\exa\electronvolt}$ can be rejected with more than $4\sigma$ confidence. At this energy, the observed proton flux produced by the second population becomes coincident with the protons from the default sources (primary or from disintegration). We thus obtain no improvement of the fit compared to the single-population scenario.

In addition, the fit is asymptotically insensitive to the maximum proton energy for $\emax^\ptwo\gtrsim E_\text{GZK}$. This justifies our choice of $10^{23}\,\si{\electronvolt}$ for the 2SC-dip scenario as a representative case for extremely-UHE proton sources.

\section{Proton Source Parameter Space}\label{apx:proton_parspace}
\begin{figure}
    \centering
    \includegraphics[width=0.48\linewidth]{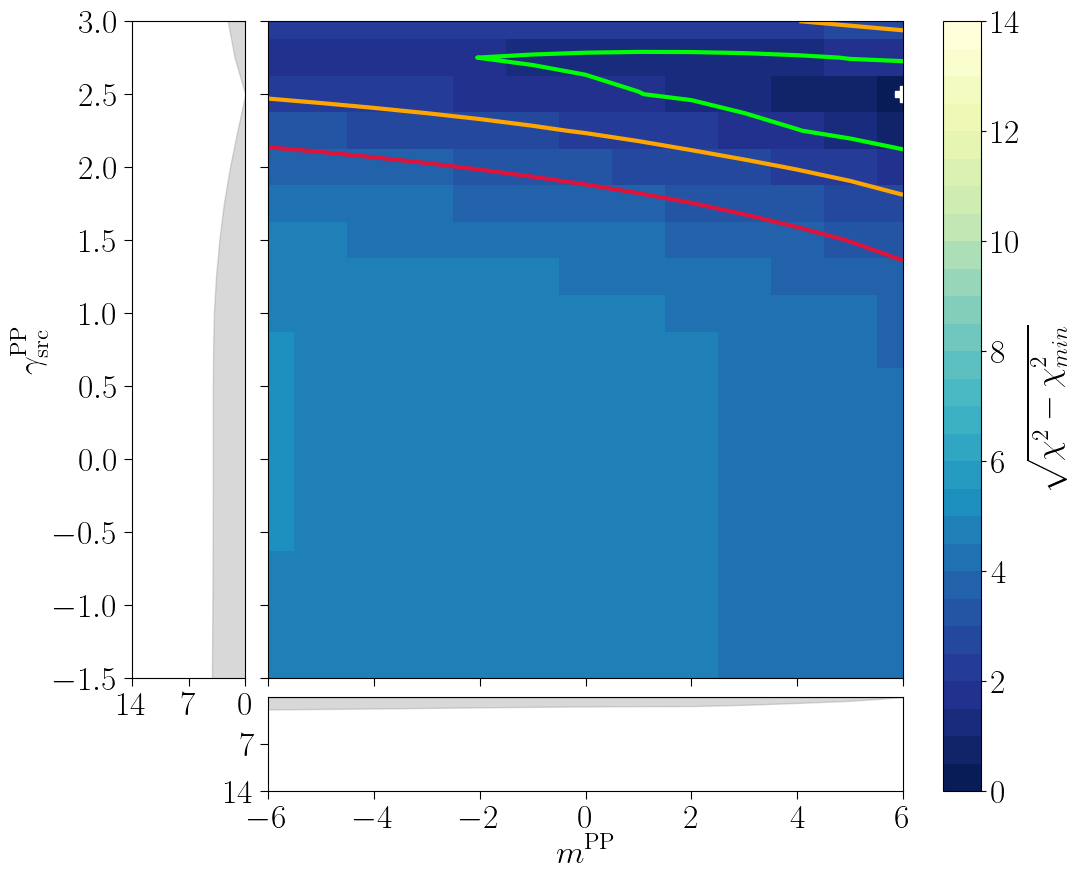}
    \includegraphics[width=0.48\linewidth]{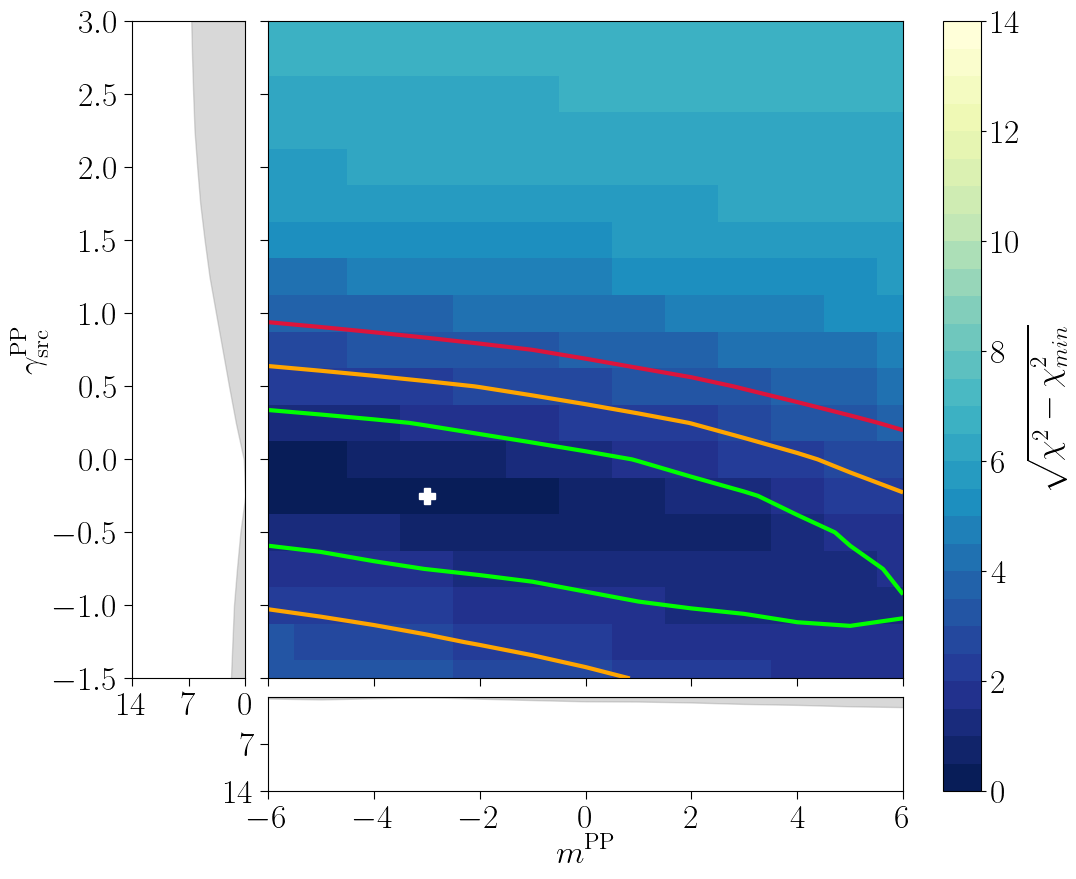}
    \caption{Fit quality for the 2SC-dip (left) and 2SC-uhecr (right) model, marginalised onto $\gamma_\text{src}^\ptwo\times m^\ptwo$ space. The best fit is marked with a white cross and contour lines indicate the one (green), two (orange) and three (red) sigma confidence intervals.}\label{fig:apx_parspace}
\end{figure}

Results of the source-parameter scan are shown in \cref{fig:apx_parspace} as a 2D surface plot over the PP injected spectral index $\gamma_\text{src}^\ptwo$ and redshift evolution $m^\ptwo$ for both the 2SC-dip and 2SC-uhecr models.


\newpage
\bibliographystyle{JHEP}
\bibliography{bibliography.bib}

\providecommand{\href}[2]{#2}\begingroup\raggedright\begin{thebibliography}{10}

\bibitem{Unger:2015laa}
M.~Unger, G.R.~Farrar and L.A.~Anchordoqui, \emph{{Origin of the ankle in the
  ultrahigh energy cosmic ray spectrum, and of the extragalactic protons below
  it}}, \href{https://doi.org/10.1103/PhysRevD.92.123001}{\emph{Phys. Rev. D}
  {\bfseries 92} (2015) 123001}
  [\href{https://arxiv.org/abs/1505.02153}{{\ttfamily 1505.02153}}].

\bibitem{Aab:2016zth}
{\scshape Pierre Auger} collaboration, \emph{{Combined fit of spectrum and
  composition data as measured by the Pierre Auger Observatory}},
  \href{https://doi.org/10.1088/1475-7516/2017/04/038}{\emph{JCAP} {\bfseries
  04} (2017) 038} [\href{https://arxiv.org/abs/1612.07155}{{\ttfamily
  1612.07155}}].

\bibitem{AlvesBatista:2018zui}
R.~Alves~Batista, R.M.~de~Almeida, B.~Lago and K.~Kotera, \emph{{Cosmogenic
  photon and neutrino fluxes in the Auger era}},
  \href{https://doi.org/10.1088/1475-7516/2019/01/002}{\emph{JCAP} {\bfseries
  01} (2019) 002} [\href{https://arxiv.org/abs/1806.10879}{{\ttfamily
  1806.10879}}].

\bibitem{Heinze:2019jou}
J.~Heinze, A.~Fedynitch, D.~Boncioli and W.~Winter, \emph{{A new view on Auger
  data and cosmogenic neutrinos in light of different nuclear disintegration
  and air-shower models}},
  \href{https://doi.org/10.3847/1538-4357/ab05ce}{\emph{Astrophys. J.}
  {\bfseries 873} (2019) 88}
  [\href{https://arxiv.org/abs/1901.03338}{{\ttfamily 1901.03338}}].

\bibitem{Ehlert:2022jmy}
D.~Ehlert, F.~Oikonomou and M.~Unger, \emph{{Curious case of the maximum
  rigidity distribution of cosmic-ray accelerators}},
  \href{https://doi.org/10.1103/PhysRevD.107.103045}{\emph{Phys. Rev. D}
  {\bfseries 107} (2023) 103045}
  [\href{https://arxiv.org/abs/2207.10691}{{\ttfamily 2207.10691}}].

\bibitem{Heinze:2020zqb}
J.~Heinze, D.~Biehl, A.~Fedynitch, D.~Boncioli, A.~Rudolph and W.~Winter,
  \emph{{Systematic parameter space study for the UHECR origin from GRBs in
  models with multiple internal shocks}},
  \href{https://doi.org/10.1093/mnras/staa2751}{\emph{Mon. Not. Roy. Astron.
  Soc.} {\bfseries 498} (2020) 5990}
  [\href{https://arxiv.org/abs/2006.14301}{{\ttfamily 2006.14301}}].

\bibitem{Peters:1961}
B.~Peters, \emph{Primary cosmic radiation and extensive air showers},
  \href{https://doi.org/10.1007/BF02783106}{\emph{Il Nuovo Cimento} {\bfseries
  22} (1961) 800}.

\bibitem{Gaisser:2016uoy}
T.K.~Gaisser, R.~Engel and E.~Resconi, \emph{{Cosmic Rays and Particle
  Physics}: {2nd Edition}}, Cambridge University Press (6, 2016).

\bibitem{Rodrigues:2020pli}
X.~Rodrigues, J.~Heinze, A.~Palladino, A.~van Vliet and W.~Winter,
  \emph{{Active Galactic Nuclei Jets as the Origin of Ultrahigh-Energy Cosmic
  Rays and Perspectives for the Detection of Astrophysical Source Neutrinos at
  EeV Energies}},
  \href{https://doi.org/10.1103/PhysRevLett.126.191101}{\emph{Phys. Rev. Lett.}
  {\bfseries 126} (2021) 191101}
  [\href{https://arxiv.org/abs/2003.08392}{{\ttfamily 2003.08392}}].

\bibitem{Waxman:1995vg}
E.~Waxman, \emph{{Cosmological gamma-ray bursts and the highest energy cosmic
  rays}}, \href{https://doi.org/10.1103/PhysRevLett.75.386}{\emph{Phys. Rev.
  Lett.} {\bfseries 75} (1995) 386}
  [\href{https://arxiv.org/abs/astro-ph/9505082}{{\ttfamily
  astro-ph/9505082}}].

\bibitem{Ishiwata:2019aet}
K.~Ishiwata, O.~Macias, S.~Ando and M.~Arimoto, \emph{{Probing heavy dark
  matter decays with multi-messenger astrophysical data}},
  \href{https://doi.org/10.1088/1475-7516/2020/01/003}{\emph{JCAP} {\bfseries
  01} (2020) 003} [\href{https://arxiv.org/abs/1907.11671}{{\ttfamily
  1907.11671}}].

\bibitem{Das:2023wtk}
S.~Das, K.~Murase and T.~Fujii, \emph{{Revisiting ultrahigh-energy constraints
  on decaying superheavy dark matter}},
  \href{https://doi.org/10.1103/PhysRevD.107.103013}{\emph{Phys. Rev. D}
  {\bfseries 107} (2023) 103013}
  [\href{https://arxiv.org/abs/2302.02993}{{\ttfamily 2302.02993}}].

\bibitem{PierreAuger:2017tlx}
{\scshape Pierre Auger} collaboration, \emph{{Inferences on mass composition
  and tests of hadronic interactions from 0.3 to 100 EeV using the
  water-Cherenkov detectors of the Pierre Auger Observatory}},
  \href{https://doi.org/10.1103/PhysRevD.96.122003}{\emph{Phys. Rev. D}
  {\bfseries 96} (2017) 122003}
  [\href{https://arxiv.org/abs/1710.07249}{{\ttfamily 1710.07249}}].

\bibitem{ToderoPeixoto:2019bt}
C.J.~Todero~Peixoto, \emph{{Estimating the Depth of Shower Maximum using the
  Surface Detectors of the Pierre Auger Observatory}},
  \href{https://doi.org/10.22323/1.358.0440}{\emph{PoS} {\bfseries ICRC2019}
  (2019) 440}.

\bibitem{Plotko:2022urd}
P.~Plotko, A.~van Vliet, X.~Rodrigues and W.~Winter, \emph{{Indication of a
  Local Source of Ultra-High-Energy Cosmic Rays in the Northern Hemisphere}},
  \href{https://arxiv.org/abs/2208.12274}{{\ttfamily 2208.12274}}.

\bibitem{Mollerach:2020mhr}
S.~Mollerach and E.~Roulet, \emph{{Extragalactic cosmic rays diffusing from two
  populations of sources}},
  \href{https://doi.org/10.1103/PhysRevD.101.103024}{\emph{Phys. Rev. D}
  {\bfseries 101} (2020) 103024}
  [\href{https://arxiv.org/abs/2004.04253}{{\ttfamily 2004.04253}}].

\bibitem{PierreAuger:2021mmt}
{\scshape Pierre Auger} collaboration, \emph{{Combined fit of the energy
  spectrum and mass composition across the ankle with the data measured at the
  Pierre Auger Observatory}},
  \href{https://doi.org/10.22323/1.395.0311}{\emph{PoS} {\bfseries ICRC2021}
  (2021) 311}.

\bibitem{Luce:2022awd}
Q.~Luce, S.~Marafico, J.~Biteau, A.~Condorelli and O.~Deligny,
  \emph{{Observational Constraints on Cosmic-Ray Escape from Ultrahigh-energy
  Accelerators}},
  \href{https://doi.org/10.3847/1538-4357/ac81cc}{\emph{Astrophys. J.}
  {\bfseries 936} (2022) 62}
  [\href{https://arxiv.org/abs/2207.08092}{{\ttfamily 2207.08092}}].

\bibitem{PierreAuger:2022atd}
{\scshape Pierre Auger} collaboration, \emph{{Constraining the sources of
  ultra-high-energy cosmic rays across and above the ankle with the spectrum
  and composition data measured at the Pierre Auger Observatory}},
  \href{https://doi.org/10.1088/1475-7516/2023/05/024}{\emph{JCAP} {\bfseries
  05} (2023) 024} [\href{https://arxiv.org/abs/2211.02857}{{\ttfamily
  2211.02857}}].

\bibitem{Muzio:2019leu}
M.S.~Muzio, M.~Unger and G.R.~Farrar, \emph{{Progress towards characterizing
  ultrahigh energy cosmic ray sources}},
  \href{https://doi.org/10.1103/PhysRevD.100.103008}{\emph{Phys. Rev. D}
  {\bfseries 100} (2019) 103008}
  [\href{https://arxiv.org/abs/1906.06233}{{\ttfamily 1906.06233}}].

\bibitem{Das:2020nvx}
S.~Das, S.~Razzaque and N.~Gupta, \emph{{Modeling the spectrum and composition
  of ultrahigh-energy cosmic rays with two populations of extragalactic
  sources}}, \href{https://doi.org/10.1140/epjc/s10052-021-08885-4}{\emph{Eur.
  Phys. J. C} {\bfseries 81} (2021) 59}
  [\href{https://arxiv.org/abs/2004.07621}{{\ttfamily 2004.07621}}].

\bibitem{Muzio:2023skc}
M.S.~Muzio, M.~Unger and S.~Wissel, \emph{{Prospects for joint cosmic ray and
  neutrino constraints on the evolution of trans-Greisen-Zatsepin-Kuzmin proton
  sources}}, \href{https://doi.org/10.1103/PhysRevD.107.103030}{\emph{Phys.
  Rev. D} {\bfseries 107} (2023) 103030}
  [\href{https://arxiv.org/abs/2303.04170}{{\ttfamily 2303.04170}}].

\bibitem{Greisen:1966jv}
K.~Greisen, \emph{{End to the cosmic ray spectrum?}},
  \href{https://doi.org/10.1103/PhysRevLett.16.748}{\emph{Phys. Rev. Lett.}
  {\bfseries 16} (1966) 748}.

\bibitem{Zatsepin:1966jv}
G.T.~Zatsepin and V.A.~Kuzmin, \emph{{Upper limit of the spectrum of cosmic
  rays}}, {\emph{JETP Lett.} {\bfseries 4} (1966) 78}.

\bibitem{Batista:2016yrx}
R.~Alves~Batista, A.~Dundovic, M.~Erdmann, K.-H.~Kampert, D.~Kuempel,
  G.~M\"uller et~al., \emph{{CRPropa 3 - a Public Astrophysical Simulation
  Framework for Propagating Extraterrestrial Ultra-High Energy Particles}},
  \href{https://doi.org/10.1088/1475-7516/2016/05/038}{\emph{JCAP} {\bfseries
  05} (2016) 038} [\href{https://arxiv.org/abs/1603.07142}{{\ttfamily
  1603.07142}}].

\bibitem{AlvesBatista:2022vem}
R.~Alves~Batista et~al., \emph{{CRPropa 3.2 \textemdash{} an advanced framework
  for high-energy particle propagation in extragalactic and galactic spaces}},
  \href{https://doi.org/10.1088/1475-7516/2022/09/035}{\emph{JCAP} {\bfseries
  09} (2022) 035} [\href{https://arxiv.org/abs/2208.00107}{{\ttfamily
  2208.00107}}].

\bibitem{Gilmore:2012}
R.C.~Gilmore, R.S.~Somerville, J.R.~Primack and A.~Domínguez,
  \emph{{Semi-analytic modelling of the extragalactic background light and
  consequences for extragalactic gamma-ray spectra}},
  \href{https://doi.org/10.1111/j.1365-2966.2012.20841.x}{\emph{Mon. Not. Roy.
  Astron. Soc.} {\bfseries 422} (2012) 3189}.

\bibitem{vanVliet:2019nse}
A.~van Vliet, R.~Alves~Batista and J.R.~H\"orandel, \emph{{Determining the
  fraction of cosmic-ray protons at ultrahigh energies with cosmogenic
  neutrinos}}, \href{https://doi.org/10.1103/PhysRevD.100.021302}{\emph{Phys.
  Rev. D} {\bfseries 100} (2019) 021302}
  [\href{https://arxiv.org/abs/1901.01899}{{\ttfamily 1901.01899}}].

\bibitem{PierreAuger:2020qqz}
{\scshape Pierre Auger} collaboration, \emph{{Measurement of the cosmic-ray
  energy spectrum above $2.5{\times} 10^{18}$ eV using the Pierre Auger
  Observatory}}, \href{https://doi.org/10.1103/PhysRevD.102.062005}{\emph{Phys.
  Rev. D} {\bfseries 102} (2020) 062005}
  [\href{https://arxiv.org/abs/2008.06486}{{\ttfamily 2008.06486}}].

\bibitem{Yushkov:2020nhr}
{\scshape Auger} collaboration, \emph{{Mass Composition of Cosmic Rays with
  Energies above 10$^{17.2}$ eV from the Hybrid Data of the Pierre Auger
  Observatory}}, \href{https://doi.org/10.22323/1.358.0482}{\emph{PoS}
  {\bfseries ICRC2019} (2020) 482}.

\bibitem{Pierog:2015}
T.~Pierog, I.~Karpenko, J.M.~Katzy, E.~Yatsenko and K.~Werner, \emph{Epos lhc:
  Test of collective hadronization with data measured at the cern large hadron
  collider}, \href{https://doi.org/10.1103/PhysRevC.92.034906}{\emph{Phys. Rev.
  C} {\bfseries 92} (2015) 034906}.

\bibitem{Fedynitch:2018cbl}
A.~Fedynitch, F.~Riehn, R.~Engel, T.K.~Gaisser and T.~Stanev, \emph{{Hadronic
  interaction model sibyll 2.3c and inclusive lepton fluxes}},
  \href{https://doi.org/10.1103/PhysRevD.100.103018}{\emph{Phys. Rev. D}
  {\bfseries 100} (2019) 103018}
  [\href{https://arxiv.org/abs/1806.04140}{{\ttfamily 1806.04140}}].

\bibitem{Fermi-LAT:2014ryh}
{\scshape Fermi-LAT} collaboration, \emph{{The spectrum of isotropic diffuse
  gamma-ray emission between 100 MeV and 820 GeV}},
  \href{https://doi.org/10.1088/0004-637X/799/1/86}{\emph{Astrophys. J.}
  {\bfseries 799} (2015) 86} [\href{https://arxiv.org/abs/1410.3696}{{\ttfamily
  1410.3696}}].

\bibitem{IceCube:2020wum}
{\scshape IceCube} collaboration, \emph{{The IceCube high-energy starting event
  sample: Description and flux characterization with 7.5 years of data}},
  \href{https://doi.org/10.1103/PhysRevD.104.022002}{\emph{Phys. Rev. D}
  {\bfseries 104} (2021) 022002}
  [\href{https://arxiv.org/abs/2011.03545}{{\ttfamily 2011.03545}}].

\bibitem{Pedreira:2021gcl}
{\scshape Pierre Auger} collaboration, \emph{{Bounds on diffuse and point
  source fluxes of ultra-high energy neutrinos with the Pierre Auger
  Observatory}}, \href{https://doi.org/10.22323/1.358.0979}{\emph{PoS}
  {\bfseries ICRC2019} (2021) 979}.

\bibitem{Savina:2021cva}
{\scshape Pierre Auger} collaboration, \emph{{A search for ultra-high-energy
  photons at the Pierre Auger Observatory exploiting air-shower universality}},
  \href{https://doi.org/10.22323/1.395.0373}{\emph{PoS} {\bfseries ICRC2021}
  (2021) 373}.

\bibitem{PierreAuger:2022aty}
{\scshape Pierre Auger} collaboration, \emph{{Search for photons above
  10$^{19}$ eV with the surface detector of the Pierre Auger Observatory}},
  \href{https://doi.org/10.1088/1475-7516/2023/05/021}{\emph{JCAP} {\bfseries
  05} (2023) 021} [\href{https://arxiv.org/abs/2209.05926}{{\ttfamily
  2209.05926}}].

\bibitem{Baker:1983tu}
S.~Baker and R.D.~Cousins, \emph{{Clarification of the Use of Chi Square and
  Likelihood Functions in Fits to Histograms}},
  \href{https://doi.org/10.1016/0167-5087(84)90016-4}{\emph{Nucl. Instrum.
  Meth.} {\bfseries 221} (1984) 437}.

\bibitem{Berezinsky:1988wi}
V.S.~Berezinsky and S.I.~Grigor'eva, \emph{{A Bump in the ultrahigh-energy
  cosmic ray spectrum}}, {\emph{Astron. Astrophys.} {\bfseries 199} (1988) 1}.

\bibitem{Heinze:2015hhp}
J.~Heinze, D.~Boncioli, M.~Bustamante and W.~Winter, \emph{{Cosmogenic
  Neutrinos Challenge the Cosmic Ray Proton Dip Model}},
  \href{https://doi.org/10.3847/0004-637X/825/2/122}{\emph{Astrophys. J.}
  {\bfseries 825} (2016) 122}
  [\href{https://arxiv.org/abs/1512.05988}{{\ttfamily 1512.05988}}].

\bibitem{Rosenfeld:1975fy}
A.H.~Rosenfeld, \emph{{The particle data group: growth and operations-eighteen
  years of particle physics}},
  \href{https://doi.org/10.1146/annurev.ns.25.120175.003011}{\emph{Ann. Rev.
  Nucl. Part. Sci.} {\bfseries 25} (1975) 555}.

\bibitem{Lisanti:2016jub}
M.~Lisanti, S.~Mishra-Sharma, L.~Necib and B.R.~Safdi, \emph{{Deciphering
  Contributions to the Extragalactic Gamma-Ray Background from 2 GeV to 2
  TeV}}, \href{https://doi.org/10.3847/0004-637X/832/2/117}{\emph{Astrophys.
  J.} {\bfseries 832} (2016) 117}
  [\href{https://arxiv.org/abs/1606.04101}{{\ttfamily 1606.04101}}].

\bibitem{HAWC:2022uka}
{\scshape HAWC} collaboration, \emph{{Limits on the Diffuse Gamma-Ray
  Background above 10 TeV with HAWC}},
  \href{https://arxiv.org/abs/2209.08106}{{\ttfamily 2209.08106}}.

\bibitem{TelescopeArray:2018rbt}
{\scshape Telescope Array} collaboration, \emph{{Constraints on the diffuse
  photon flux with energies above $10^{18}$ eV using the surface detector of
  the Telescope Array experiment}},
  \href{https://doi.org/10.1016/j.astropartphys.2019.03.003}{\emph{Astropart.
  Phys.} {\bfseries 110} (2019) 8}
  [\href{https://arxiv.org/abs/1811.03920}{{\ttfamily 1811.03920}}].

\bibitem{GRAND:2018iaj}
{\scshape GRAND} collaboration, \emph{{The Giant Radio Array for Neutrino
  Detection (GRAND): Science and Design}},
  \href{https://doi.org/10.1007/s11433-018-9385-7}{\emph{Sci. China Phys. Mech.
  Astron.} {\bfseries 63} (2020) 219501}
  [\href{https://arxiv.org/abs/1810.09994}{{\ttfamily 1810.09994}}].

\bibitem{IceCube:2018fhm}
{\scshape IceCube} collaboration, \emph{{Differential limit on the
  extremely-high-energy cosmic neutrino flux in the presence of astrophysical
  background from nine years of IceCube data}},
  \href{https://doi.org/10.1103/PhysRevD.98.062003}{\emph{Phys. Rev. D}
  {\bfseries 98} (2018) 062003}
  [\href{https://arxiv.org/abs/1807.01820}{{\ttfamily 1807.01820}}].

\bibitem{PierreAuger:2019ens}
{\scshape Pierre Auger} collaboration, \emph{{Probing the origin of
  ultra-high-energy cosmic rays with neutrinos in the EeV energy range using
  the Pierre Auger Observatory}},
  \href{https://doi.org/10.1088/1475-7516/2019/10/022}{\emph{JCAP} {\bfseries
  10} (2019) 022} [\href{https://arxiv.org/abs/1906.07422}{{\ttfamily
  1906.07422}}].

\bibitem{IceCube:2019pna}
{\scshape IceCube} collaboration, M.G.~Aartsen et~al., \emph{{Neutrino
  astronomy with the next generation IceCube Neutrino Observatory}},  11, 2019.

\bibitem{ARA:2015wxq}
{\scshape ARA} collaboration, \emph{{Performance of two Askaryan Radio Array
  stations and first results in the search for ultrahigh energy neutrinos}},
  \href{https://doi.org/10.1103/PhysRevD.93.082003}{\emph{Phys. Rev. D}
  {\bfseries 93} (2016) 082003}
  [\href{https://arxiv.org/abs/1507.08991}{{\ttfamily 1507.08991}}].

\bibitem{Cummings:2020ycz}
A.L.~Cummings, R.~Aloisio and J.F.~Krizmanic, \emph{{Modeling of the Tau and
  Muon Neutrino-induced Optical Cherenkov Signals from Upward-moving Extensive
  Air Showers}}, \href{https://doi.org/10.1103/PhysRevD.103.043017}{\emph{Phys.
  Rev. D} {\bfseries 103} (2021) 043017}
  [\href{https://arxiv.org/abs/2011.09869}{{\ttfamily 2011.09869}}].

\bibitem{Coleman:2022abf}
A.~Coleman et~al., \emph{{Ultra high energy cosmic rays The intersection of the
  Cosmic and Energy Frontiers}},
  \href{https://doi.org/10.1016/j.astropartphys.2023.102819}{\emph{Astropart.
  Phys.} {\bfseries 149} (2023) 102819}
  [\href{https://arxiv.org/abs/2205.05845}{{\ttfamily 2205.05845}}].

\bibitem{PUEO:2020bnn}
{\scshape PUEO} collaboration, \emph{{The Payload for Ultrahigh Energy
  Observations (PUEO): a white paper}},
  \href{https://doi.org/10.1088/1748-0221/16/08/P08035}{\emph{JINST} {\bfseries
  16} (2021) P08035} [\href{https://arxiv.org/abs/2010.02892}{{\ttfamily
  2010.02892}}].

\bibitem{Aloisio:2000cm}
R.~Aloisio, P.~Blasi, P.L.~Ghia and A.F.~Grillo, \emph{{Probing the structure
  of space-time with cosmic rays}},
  \href{https://doi.org/10.1103/PhysRevD.62.053010}{\emph{Phys. Rev. D}
  {\bfseries 62} (2000) 053010}
  [\href{https://arxiv.org/abs/astro-ph/0001258}{{\ttfamily
  astro-ph/0001258}}].

\bibitem{Scully:2008jp}
S.T.~Scully and F.W.~Stecker, \emph{{Lorentz Invariance Violation and the
  Observed Spectrum of Ultrahigh Energy Cosmic Rays}},
  \href{https://doi.org/10.1016/j.astropartphys.2009.01.002}{\emph{Astropart.
  Phys.} {\bfseries 31} (2009) 220}
  [\href{https://arxiv.org/abs/0811.2230}{{\ttfamily 0811.2230}}].

\bibitem{PierreAuger:2021tog}
{\scshape Pierre Auger} collaboration, \emph{{Testing effects of Lorentz
  invariance violation in the propagation of astroparticles with the Pierre
  Auger Observatory}},
  \href{https://doi.org/10.1088/1475-7516/2022/01/023}{\emph{JCAP} {\bfseries
  01} (2022) 023} [\href{https://arxiv.org/abs/2112.06773}{{\ttfamily
  2112.06773}}].

\bibitem{Murase:2018utn}
K.~Murase and M.~Fukugita, \emph{{Energetics of High-Energy Cosmic
  Radiations}}, \href{https://doi.org/10.1103/PhysRevD.99.063012}{\emph{Phys.
  Rev. D} {\bfseries 99} (2019) 063012}
  [\href{https://arxiv.org/abs/1806.04194}{{\ttfamily 1806.04194}}].

\bibitem{Berezinsky:1997hy}
V.~Berezinsky, M.~Kachelriess and A.~Vilenkin, \emph{{Ultrahigh-energy cosmic
  rays without GZK cutoff}},
  \href{https://doi.org/10.1103/PhysRevLett.79.4302}{\emph{Phys. Rev. Lett.}
  {\bfseries 79} (1997) 4302}
  [\href{https://arxiv.org/abs/astro-ph/9708217}{{\ttfamily
  astro-ph/9708217}}].

\bibitem{Kuzmin:1997jua}
V.A.~Kuzmin and V.A.~Rubakov, \emph{{Ultrahigh-energy cosmic rays: A Window to
  postinflationary reheating epoch of the universe?}}, {\emph{Phys. Atom.
  Nucl.} {\bfseries 61} (1998) 1028}
  [\href{https://arxiv.org/abs/astro-ph/9709187}{{\ttfamily
  astro-ph/9709187}}].

\bibitem{Sigl:1998vz}
G.~Sigl, S.~Lee, P.~Bhattacharjee and S.~Yoshida, \emph{{Probing grand unified
  theories with cosmic ray, gamma-ray and neutrino astrophysics}},
  \href{https://doi.org/10.1103/PhysRevD.59.043504}{\emph{Phys. Rev. D}
  {\bfseries 59} (1999) 043504}
  [\href{https://arxiv.org/abs/hep-ph/9809242}{{\ttfamily hep-ph/9809242}}].

\bibitem{Bhattacharjee:1999mup}
P.~Bhattacharjee and G.~Sigl, \emph{{Origin and propagation of extremely
  high-energy cosmic rays}},
  \href{https://doi.org/10.1016/S0370-1573(99)00101-5}{\emph{Phys. Rept.}
  {\bfseries 327} (2000) 109}
  [\href{https://arxiv.org/abs/astro-ph/9811011}{{\ttfamily
  astro-ph/9811011}}].

\bibitem{Ellis:2005jc}
J.R.~Ellis, V.E.~Mayes and D.V.~Nanopoulos, \emph{{Uhecr particle spectra from
  crypton decays}},
  \href{https://doi.org/10.1103/PhysRevD.74.115003}{\emph{Phys. Rev. D}
  {\bfseries 74} (2006) 115003}
  [\href{https://arxiv.org/abs/astro-ph/0512303}{{\ttfamily
  astro-ph/0512303}}].

\bibitem{Kalashev:2008dh}
O.E.~Kalashev, G.I.~Rubtsov and S.V.~Troitsky, \emph{{Sensitivity of cosmic-ray
  experiments to ultra-high-energy photons: reconstruction of the spectrum and
  limits on the superheavy dark matter}},
  \href{https://doi.org/10.1103/PhysRevD.80.103006}{\emph{Phys. Rev. D}
  {\bfseries 80} (2009) 103006}
  [\href{https://arxiv.org/abs/0812.1020}{{\ttfamily 0812.1020}}].

\bibitem{Aloisio:2015lva}
R.~Aloisio, S.~Matarrese and A.V.~Olinto, \emph{{Super Heavy Dark Matter in
  light of BICEP2, Planck and Ultra High Energy Cosmic Rays Observations}},
  \href{https://doi.org/10.1088/1475-7516/2015/08/024}{\emph{JCAP} {\bfseries
  08} (2015) 024} [\href{https://arxiv.org/abs/1504.01319}{{\ttfamily
  1504.01319}}].

\bibitem{Supanitsky:2019ayx}
A.D.~Supanitsky and G.~Medina-Tanco, \emph{{Ultra high energy cosmic rays from
  super-heavy dark matter in the context of large exposure observatories}},
  \href{https://doi.org/10.1088/1475-7516/2019/11/036}{\emph{JCAP} {\bfseries
  11} (2019) 036} [\href{https://arxiv.org/abs/1909.09191}{{\ttfamily
  1909.09191}}].

\bibitem{Kofman:1994rk}
L.~Kofman, A.D.~Linde and A.A.~Starobinsky, \emph{{Reheating after inflation}},
  \href{https://doi.org/10.1103/PhysRevLett.73.3195}{\emph{Phys. Rev. Lett.}
  {\bfseries 73} (1994) 3195}
  [\href{https://arxiv.org/abs/hep-th/9405187}{{\ttfamily hep-th/9405187}}].

\bibitem{Felder:1998vq}
G.N.~Felder, L.~Kofman and A.D.~Linde, \emph{{Instant preheating}},
  \href{https://doi.org/10.1103/PhysRevD.59.123523}{\emph{Phys. Rev. D}
  {\bfseries 59} (1999) 123523}
  [\href{https://arxiv.org/abs/hep-ph/9812289}{{\ttfamily hep-ph/9812289}}].

\bibitem{Chung:1998rq}
D.J.H.~Chung, E.W.~Kolb and A.~Riotto, \emph{{Production of massive particles
  during reheating}},
  \href{https://doi.org/10.1103/PhysRevD.60.063504}{\emph{Phys. Rev. D}
  {\bfseries 60} (1999) 063504}
  [\href{https://arxiv.org/abs/hep-ph/9809453}{{\ttfamily hep-ph/9809453}}].

\bibitem{PierreAuger:2016kuz}
{\scshape Pierre Auger} collaboration, \emph{{Search for photons with energies
  above 10$^{18}$ eV using the hybrid detector of the Pierre Auger
  Observatory}},
  \href{https://doi.org/10.1088/1475-7516/2017/04/009}{\emph{JCAP} {\bfseries
  04} (2017) 009} [\href{https://arxiv.org/abs/1612.01517}{{\ttfamily
  1612.01517}}].

\bibitem{Rautenberg:2021vvt}
{\scshape Pierre Auger} collaboration, \emph{{Limits on ultra-high energy
  photons with the Pierre Auger Observatory}},
  \href{https://doi.org/10.22323/1.358.0398}{\emph{PoS} {\bfseries ICRC2019}
  (2021) 398}.

\bibitem{Kuznetsov:2016fjt}
M.Y.~Kuznetsov, \emph{{Hadronically decaying heavy dark matter and high-energy
  neutrino limits}},
  \href{https://doi.org/10.1134/S0021364017090028}{\emph{JETP Lett.} {\bfseries
  105} (2017) 561} [\href{https://arxiv.org/abs/1611.08684}{{\ttfamily
  1611.08684}}].

\bibitem{Cohen:2016uyg}
T.~Cohen, K.~Murase, N.L.~Rodd, B.R.~Safdi and Y.~Soreq,
  \emph{{\ensuremath{\gamma} -ray Constraints on Decaying Dark Matter and
  Implications for IceCube}},
  \href{https://doi.org/10.1103/PhysRevLett.119.021102}{\emph{Phys. Rev. Lett.}
  {\bfseries 119} (2017) 021102}
  [\href{https://arxiv.org/abs/1612.05638}{{\ttfamily 1612.05638}}].

\bibitem{Kachelriess:2018rty}
M.~Kachelriess, O.E.~Kalashev and M.Y.~Kuznetsov, \emph{{Heavy decaying dark
  matter and IceCube high energy neutrinos}},
  \href{https://doi.org/10.1103/PhysRevD.98.083016}{\emph{Phys. Rev. D}
  {\bfseries 98} (2018) 083016}
  [\href{https://arxiv.org/abs/1805.04500}{{\ttfamily 1805.04500}}].

\bibitem{PierreAuger:2022ubv}
{\scshape Pierre Auger} collaboration, \emph{{Cosmological implications of
  photon-flux upper limits at ultrahigh energies in scenarios of
  Planckian-interacting massive particles for dark matter}},
  \href{https://doi.org/10.1103/PhysRevD.107.042002}{\emph{Phys. Rev. D}
  {\bfseries 107} (2023) 042002}
  [\href{https://arxiv.org/abs/2208.02353}{{\ttfamily 2208.02353}}].

\end{thebibliography}\endgroup

\end{document}